\documentclass[10pt]{article}
\usepackage{times}
\usepackage{graphicx}
\usepackage{amssymb}
\usepackage{amsmath}
\usepackage{multirow}
\usepackage[square,numbers,sort&compress]{natbib}
\newcommand{\bm}[1]{\mbox{\boldmath${#1}$}}
\newcommand{\lw}[1]{\smash{\lower1.5ex\hbox{#1}}}
\newcommand{\Slash}[1]{{/\hspace{-0.23cm}{#1}}}
\newcommand{\ds}[1]{\displaystyle{#1}}
\newcommand{\beq}{\begin{eqnarray}}
\newcommand{\eeq}{\end{eqnarray}}
\newcommand{\nn}{\nonumber}
\newcommand{\bra}{\langle}
\newcommand{\ket}{\rangle}
\newcommand{\del}{\partial}

\newcommand{\diag}{\mbox{diag}}

\def\calL{{\cal L}}    
    
\newcommand{\half}{\frac{1}{2}}
\newcommand{\thalf}{\frac{3}{2}}

\newcommand{\inner}[2]{{\bm #1}\cdot {\bm #2}}

\def\Iprot#1#2{P^{#1#2}_{3/2}}

\def\Sprot#1#2{\Gamma^{#1#2}_{3/2}}

\def\Hcterms{({\rm H.c.})}
\def\nrm{ { \rm n}}
\def\addriise{{\it Research Institute for Information Science and Education, Hiroshima University,}\\
{\it Higashi-Hiroshima 739-8527 JAPAN} \\
{\it Present address : Department of Physics, The University of Tokyo,} \\
{\it Bunkyo-ku, Tokyo 113-0033 JAPAN }}
\begin{document}
\title{Quartet of spin-3/2 baryons in chiral multiplet $(1, 1/2) \oplus (1/2, 1) $ with mirror assignment}
\author{Keitaro Nagata \\ \addriise}
\date{\today}
%
\maketitle
We study the possible existence of chiral partners in the spin-$\thalf$ sector of the
baryon spectrum. We consider a quartet scheme where four spin-3/2 baryons, $P_{33}$, 
$D_{33}$, $D_{13}$ and $P_{13}$, group into higher-dimensional chiral multiplets 
$(1, \half)\oplus (\half,1)$ with a mirror assignment. With an effective 
$SU(2)_R\times SU(2)_L$ Lagrangian, we derive constraints imposed by chiral 
symmetry together with the mirror assignment on the masses and coupling 
constants of the quartet. Using the effective Lagrangian, we try to find a set of 
baryons suitable for the chiral quartet.
It turns out that two cases reasonably agree with the mass pattern of the quartet:
($\Delta(1600)$, $\Delta(1940)$, $N(1520)$, $N(1720)$) and  
($\Delta(1920)$, $\Delta(1940)$, $N(2080)$, $N(1900)$).

\section{Introduction}

Chiral symmetry $SU(N_F)_R\times SU(N_F)_L$ and its spontaneous breaking 
characterize the QCD vacuum, and is a key to understanding the strong interactions. 
Due to the spontaneous breaking of chiral symmetry (SBCS), the hadron spectrum is 
classified in terms of the residual symmetry $SU(N_F)_V$, while the role of 
$SU(N_F)_R \times SU(N_F)_L$ in the hadron spectrum is unclear. Nevertheless, one 
expects that there exists a set of hadrons reflecting a nature of the original 
symmetry, which is referred to as chiral partners. Such examples are well-known 
for mesons, e.g. $(\sigma, \pi)$ and $(\rho, a_1)$~\cite{Vogl:1991qt,Klevansky:1992qe,Hatsuda:1994pi}, 
while not well established for baryons. As discussed in the meson's case, 
finding chiral partners provides us with the understanding of the role of 
chiral symmetry in the hadron spectrum, and also a clue to study the restoration 
of chiral symmetry. Recently, the multiplet nature of the chiral group draws a 
renewed attention from an interest in the effective chiral 
restoration~\cite{Cohen:2006bq,Glozman:2007jt,Glozman:2007ek}, which was 
suggested to be the cause of the observed parity doubling in high-energy region 
of the spectrum~\cite{Jaffe:2006jy}.

In the present work, we address the issue of the multiplet nature of the baryon's chiral 
partners. We denote a chiral multiplet by $(I_R, I_L)$, where $I_R[I_L]$ is
an isospin for $SU(2)_R[SU(2)_L]$. All the members of one chiral multiplet $(I_R, I_L)$ 
have a fixed spin. The correspondence of the charge algebra between $SU(N_f)_R\times SU(N_f)_L$ and 
$SU(N_f)_V \times SU(N_f)_A$ leads to a relation $I = I_R\oplus I_L = | I_R + I_L| , \cdots, |I_R - I_L|$. 
This implies that a chiral multiplet can contain various isospin states. 
In the presence of the SBCS, the mixing of different chiral representations 
happens, and a hadron with an isospin $I$ can be described as a superposition of 
various chiral representations containing $I$. We are here concerned with 
the case that a set of hadrons group into one or a few representations 
even in the presence of the SBCS, or the case where the configuration mixing is small.

In order to find chiral partners, we need to understand the multiplet nature of the chiral 
group, such as the pattern of the spectrum and coupling constants of the multiplet. Because 
general relations for masses and axial charges that can be applied to arbitrary chiral 
representations are not established so far, the properties of the chiral partners are 
usually studied with focusing on a particular chiral representation. In the meson's case, 
the properties of chiral partners have been investigated by using e.g. the NJL 
model~\cite{Nambu:1961tp,Nambu:1961fr} and Weinberg sum rules~\cite{Kapusta:1993hq}.
The NJL model was applied to the nucleon~\cite{Ishii:1993xm,Ishii:1993rt,Ishii:2000zy,Mineo:2002bg,Bentz:2002um} 
and  $\Delta(1232)$~\cite{Ishii:1995bu} by solving the Faddeev equation. We applied the 
NJL model with diquarks  to the nucleon~\cite{Abu-Raddad:2002pw,Nagata:2003gg,Nagata:2004ky} 
and the Roper resonance~\cite{Nagata:2005qb}, using an auxiliary field method. However, 
when we apply such microscopic approaches to a baryon with a mass larger than sum of 
the masses of the internal degrees of freedom, we encounter the difficulty of the confinement.
Due to this difficulty, effective Lagrangian approaches that contain hadrons as degrees of 
freedom are often employed for the study of baryon's chiral partners~\cite{Gell-Mann:1960np,Lee:1973,DeTar:1988kn,Jido:1998av,Jido:2001nt,Nagahiro:2008rj,Gallas:2009qp,Sasaki:2010bp}.

In recent papers, we have developed a systematic method to construct an effective 
$SU(N_f)_R\times SU(N_F)_L$ Lagrangian including higher-dimensional representations~\cite{Nagata:2007di,Nagata:2008xf,Chen:2008qv,Nagata:2008cq,Dmitrasinovic:2009vp}, 
which we refer to as a projection method. This method is inspired by an 
NJL model for mesons, and partly extend it to baryons. In Ref.~\cite{Nagata:2007di}, 
we classified baryon fields consisting of three quarks in terms of chiral multiplets. 
The Pauli principle implemented by the Fierz transformation plays a crucial role in 
the classification. The projection method is performed as follows. First we find 
a chiral invariant operator involving direct products of the quark and diquark 
fields. This can be achieved by using  an analogy between $(\sigma, \vec{\pi})$ 
and diquarks in chiral transformation property. Then, we project the direct products 
of the quark and diquark fields onto irreducible parts with the use of the Fierz identities.
After the projection, three-quark fields are replaced by baryon fields.
Thus we can systematically construct chiral invariant Lagrangians including higher-dimensional 
chiral representations, avoiding problems caused by the lack of the confinement. Although such 
simple effective Lagrangians have limited validity, they are useful for the present purpose to
derive the pattern of the masses and coupling constants of the chiral multiplet. 

In Ref.~\cite{Nagata:2008xf}, we have applied the projection method to a quartet 
scheme (QS). The QS was first proposed by Jido et. al.~\cite{Jido:1999hd}. 
They used two kinds of $(1,\half)\oplus (\half,1)$ and considered so-called mirror 
assignment~\cite{Lee:1973,DeTar:1988kn,Jido:2001nt}, where  four types of 
baryons, two with $I=\half$ and the other two with $I=\thalf$, are included in the 
multiplet. They applied the QS to $J=\half, \thalf$ and $\frac{5}{2}$ and studied the 
masses and intra-coupling constants of the quartet. They did not consider Dirac structure 
of the Lagrangian explicitly. Owing to the projection method, we took intro account 
the Dirac structure in the QS Lagrangian which enables us to include transition terms 
between $J=\half$ and $J=\thalf$, e.g. $N$ and $\Delta(1232)$. With the QS Lagrangian, 
we have derived several constraints on the masses and coupling constants, which 
characterize the multiplet nature of the quartet.

In the present work, we develop the previous study to find a set of baryons suitable for the 
chiral quartet of spin-$\thalf$ baryons. Considering $J=\thalf$, the quartet consists of 
$P_{33}$, $D_{33}$, $D_{13}$, and $P_{13}$. Among various candidates for this set, we adopted 
a particular assignment in Ref.~\cite{Nagata:2008xf}: $\Delta(1232)$, $\Delta(1700)$, $N(1520)$, 
$N(1720)$. It is an important question if there is other assignment suitable for the quartet.
One interesting assignment is a set ($\Delta(1920)$, $\Delta(1940)$, $N(2080)$, $N(1900)$). 
Glozman mentioned the possibility that the approximate degeneracy of these four baryons is 
a consequence of the effective chiral restoration~\cite{Glozman:2007ek}. If this is the case, 
there are two possibilities. The first one is that the four baryons form the chiral quartet. 
The second one is that two $\Delta$s belong to $(\thalf, 0)\oplus (0, \thalf)$ and two 
$N^*$ belong to $(\half, 0)\oplus (0, \half)$. We can study the first case using the QS Lagrangian. 

In order to take into account $\pi N$ interactions in the QS, it is necessary to 
determine the nucleon's chiral representation. In standard linear $\sigma$ models of 
Gell-Mann-Levy type~\cite{Gell-Mann:1960np} the nucleon belongs to $(\half,0) \oplus (0, \half)$. 
In the mirror models~\cite{Lee:1973,DeTar:1988kn,Jido:1998av,Jido:2001nt,Nagahiro:2008rj,Gallas:2009qp}, 
the nucleon is a mixture of two kinds of $(\half,0)\oplus (0, \half)$. 
The mixing of $(\half,0)\oplus (0,\half)$ and $(1,\half)\oplus (\half,1)$ was studied 
in an algebraic approach~\cite{Weinberg:1969hw,Beane:2002ud,Hosaka:2001ti} and field 
theoretical approaches~\cite{Nagata:2008cq,Dmitrasinovic:2009vp}. In non-relativistic 
quark models the nucleon wave-functions also correspond to the mixing of 
$(1,\half) \oplus (\half, 1)$ and $(\half,0) \oplus (0, \half)$.  In the present study, 
we assume the nucleon to be saturated with the fundamental representation 
$(\half, 0) \oplus (0, \half)$ due to the following reasons. The linear $\sigma$ models 
qualitatively describe the chiral properties of the nucleon. For instance, the linear 
$\sigma$ models describe $g_A=1$ in qualitative agreement with 
$g_A^{\rm (exp)}= 1.267\pm 0.004$. Secondly, the nucleon belongs to $(\half, 0)\oplus (0, \half)$, 
if the nucleon operator has spatially symmetric property~\cite{Nagata:2007di}.

This paper is organized as follows. In section~\ref{May2509sec1}, we define 
the baryon fields and derive their $SU(2)_A$ transformation properties. 
In section~\ref{May2509sec2}, we construct the $SU(2)_R\times SU(2)_L$ Lagrangian, 
such as mass terms and $\pi NR$ interactions  with the use of the projection technique. 
Here $R$ denotes the member of the chiral quartet. Although the QS Lagrangian is 
not new, we generalize the formulation given in the previous study in a assignment-free manner 
in order make it feasible to test various assignment. With the Lagrangian, we derive 
several constraints on the properties of  the quartet. Because the projection method 
is complicated,  we show an alternative derivation of some of the present results, 
using chiral algebra in Appendix~\ref{Feb0110sec1}. Numerical results are shown 
in section~\ref{May2509sec3}. Considering the masses, we find two suitable assignments
 ($\Delta(1600)$, $\Delta(1940)$, $N(1520)$, $N(1720)$) and  
 ($\Delta(1920)$, $\Delta(1940)$, $N(2080)$, $N(1900)$). We discuss the properties of 
the quartet for these cases together with the assignment ($\Delta(1232)$, 
$\Delta(1700)$, $N(1520)$, $N(1720)$). The final section is devoted to a summary.

\section{Chiral Properties of Baryon Fields}
\label{May2509sec1}
In this section, we consider baryon fields consisting of three quarks, which 
serves as a preparation for the projection method. Baryon fields consisting 
of three quarks in a local form are generally described as
\begin{eqnarray}
B(x)\sim \epsilon_{abc} \left(q^T_a(x) \Gamma_1 q_b(x)\right) \Gamma_2 q_c(x),
\label{eq:bgeneral}
\end{eqnarray}
where $q(x)=(u(x),\;d(x))^T$ is an iso-doublet quark field at location $x$, 
the superscript $T$ represents the transpose and the indices $a,\;b$ and 
$c$ represent the color. The antisymmetric tensor in color space $\epsilon_{abc}$ 
ensures the baryons being color singlets. From now on, we shall omit 
the color indices and space-time coordinates. $\Gamma_{1,2}$ describe
 Dirac and isospin matrices. With a suitable choice of $\Gamma_{1,2}$, 
a baryon field is defined so that it forms an irreducible representation 
of the Lorentz and isospin groups.

Concerning $J=\thalf$, there are three possible baryon fields with $I=\half$;
\begin{subequations}
\begin{align}
N_{V}^{\mu} &= (\tilde{q}\gamma_\nu q) \Gamma^{\mu \nu}_{3/2} \gamma_5 q, \\
N_{A}^{\mu} &= (\tilde{q}\gamma_\nu \gamma_5 \tau^i q)  \Gamma^{\mu \nu}_{3/2}\tau^i q, \\
N_{T}^{\mu} &= i (\tilde{q} \sigma_{\alpha\beta} \tau^i q)  \Sprot{\mu}{\alpha}\gamma^\beta\gamma_5 \tau^i q, 
\end{align}
\label{eq:22feb2008eq1}
and two  with $I=\thalf$;
\begin{align}
\Delta_{A}^{\mu i} &= (\tilde{q} \gamma_\nu \gamma_5 \tau^j q)   \Gamma^{\mu \nu}_{3/2}P^{ij}_{3/2} q,\\
\Delta_{T}^{\mu i}&=i (\tilde{q} \sigma_{\alpha\beta}\tau^j q)
\Sprot{\mu}{\alpha}\gamma^\beta\gamma_5 \Iprot{i}{j} q.
\end{align}%
\label{eq:22feb2008eq2}%
\end{subequations}%
where $\tilde{q}=q^T C (i\tau_2)\gamma_5$ is a transposed quark field.
Here we employ an isospurion formalism~\cite{Hemmert:1997ye,Pascalutsa:2006up} 
for an isospin-$\thalf$ projection operator  $P^{ij}_{3/2}$, which is given by  
$P^{ij}_{3/2}=  \delta^{ij}-\frac13 \tau^i \tau^j. $
Similarly, $\Gamma^{\mu\nu}_{3/2}$ is a local spin-$\thalf$ projection
operator defined by $\Gamma^{\mu\nu}_{3/2}  =g^{\mu\nu}-\frac14 \gamma^\mu \gamma^\nu$.
In the present work, we consider only on-shell spin-$\thalf$ states. 
In order to consider  off-shell spin-$\thalf$ baryons, we need to employ the non-local projector 
instead of the local one~\cite{Benmerrouche:1989uc,Haberzettl:1998rw,Pascalutsa:1999zz,Pascalutsa:2005nd}.

Note that the baryon fields Eqs.~(\ref{eq:22feb2008eq1}) are not independent~\cite{Ioffe:1981kw,Chung:1981cc,Espriu:1983hu}.
In addition, they belong to reducible chiral representations, which leads to unphysical 
mixings of different chiral representations~\cite{Nagata:2007di}. 
The cause of the unphysical chiral mixings is  the fact that Eqs.~(\ref{eq:22feb2008eq1}) 
are not totally anti-symmetric; they are anti-symmetric only for the interchange between 
the first and second quarks. Considering the Fierz transformation as the anti-symmetrization 
of the second and third quarks, we obtain the totally-antisymmetric baryon fields
\begin{subequations}
\begin{align}
N_1^\mu &= \frac{\sqrt{3}}{4} N_V^\mu + \frac{1}{4\sqrt{3}} N_A^\mu,\\
\Delta_1^{\mu i} &= \frac{1}{2} \Delta_A^{\mu i}.
\end{align}%
\label{Sep2709eq1}%
\end{subequations}%
These totally-antisymmetric combinations belong to the irreducible chiral multiplet~\cite{Nagata:2007di}. 
The derivation of Eq.~(\ref{Sep2709eq1}) is shown  in Appendix~\ref{Oct0209sec1}.

With the baryon fields consisting of the quark fields, it is 
straightforward but tedious task to derive their $SU(2)_A$ transformations 
by using that of the quark field : $\delta_5^{\vec{a}} q = \half i \inner{a}{\tau} \gamma_5 q$ with 
$\vec{a}$ being the infinitesimal parameters for $SU(2)_A$. We obtain 
\begin{subequations}
\begin{align}
\delta_5^{\vec{a}} N_1^\mu &= \half \left( \frac53 i \inner{a}{\tau}\gamma_5
N_1^\mu + \frac{4}{\sqrt{3}}i\gamma_5\inner{a}{\Delta_1^\mu}
\right),\label{eq:chiralnmunv}  \\
\delta_5^{\vec{a}} \Delta_1^{\mu i} &=\half \left( \frac{4}{\sqrt{3}} i\gamma_5 a^j P^{ij}_{\frac32} N_1^\mu
-\frac23 i \tau^i\gamma_5\inner{a}{\Delta_1^\mu}+i\inner{a}{\tau}\gamma_5
\Delta_1^{\mu i}\right),
\label{eq:chiraldelmunv}%
\end{align}%
\label{Jul0608eq1}%
\end{subequations}%
which contain off-diagonal terms $\delta_5^{\vec{a}} N_1^\mu\sim {\Delta_1^{\mu i}}$ 
and $\delta_5^{\vec{a}} \Delta_1^{\mu i} \sim N_1^\mu$ as well as the diagonal ones. 
They restrict possible chiral invariant terms, similar to the case of $(\sigma, \pi)$ 
in the linear sigma model.

For later convenience, we define diquark fields contained in the spin-$\thalf$ baryon fields:
a Lorentz vector isoscalar diquark $V^\mu$ ($I(J)^P=0(1)^-$), a Lorentz axial-vector 
isovector diquark $A^{\mu i}$ ($1(1)^+$)
\begin{subequations}
\begin{align}
V^{\mu } &= \tilde{q}  \gamma^\mu q, \\
A^{\mu i} &=\tilde{q}  \gamma^\mu\gamma_5 \tau^i q.
\end{align}%
\label{Apr1008eq1}%
\end{subequations}%
It is easy to check that $V^{\mu}$ and $A^{\mu i}$ correspond to $\sigma$ and $\vec{\pi}$ mesons
in chiral transformation properties, which is a key of the projection method.

We introduce the other set of $(1,\half)\oplus (\half,1)$: $(N_2^\mu,\Delta_2^{\mu i})$, where 
they have the same spin and isospin as the original ones $(N_1^\mu,\Delta_1^{\mu i})$, 
but the opposite $SU(2)_A$ transformation properties in sign, i.e.,
\begin{subequations}
\begin{align}
\delta_5^{\vec{a}} N_2^\mu &=  - \half \left(\frac53 i \inner{a}{\tau}\gamma_5
N_2^\mu + \frac{4}{\sqrt{3}}i\gamma_5\inner{a}{\Delta_2^\mu}\right)
\label{eq:chiralnmumr} \\
\delta_5^{\vec{a}} \Delta_2^{\mu i}&= - \half \left(\frac{4}{\sqrt{3}} i\gamma_5 a^j 
P^{ij}_{\frac32} N_2^\mu
-\frac23 i \tau^i\gamma_5\inner{a}{\Delta_2^\mu}+i\inner{a}{\tau}\gamma_5
\Delta_2^{\mu i}\right).
\label{eq:chiraldelmumr}
\end{align}%
\label{Sep2509eq1}%
\end{subequations}%
This property is referred to as the mirror assignment~\cite{Jido:2001nt}, and 
we refer to $(N_1^\mu,\Delta_1^{\mu i})$ as naive, and to $(N_2^\mu,\Delta_2^{\mu i})$ 
as mirror. There is a correspondence of the chiral transformation properties between 
the naive and mirror sets, 
\begin{align}
(N_{1 R}^\mu, N_{1L}^\mu, \Delta_{1 R}^{\mu i}, \Delta_{1 L}^{\mu i}) \leftrightarrow
(N_{2 L}^\mu, N_{2R}^\mu, \Delta_{2 L}^{\mu i}, \Delta_{2 R}^{\mu i}),
\label{Mar0209eq2}
\end{align}
where the indices $R$ and $L$ denote the left- and right-handed projection 
with the projection operator $P_{R,L} = (1\pm \gamma_5)/2$. The right-handed 
parts of $N_1^\mu$ and $\Delta_1^{\mu i}$ have the same chiral transformation 
properties as the left-handed parts of $N_2^\mu$ and $\Delta_2^{\mu i}$, 
and vice versa. 

Note that we defined $N_2$ and $\Delta_2$ by their transformation properties 
Eqs.~(\ref{Sep2509eq1}). It is useful to define the baryon fields for $N_2$ 
and $\Delta_2$. It is impossible to describe them in terms of local three-quark 
fields. Since baryons are composite particles, there are generally various possible 
expressions for $N_2$ and $\Delta_2$. For example, we can describe them by 
using baryon operators having a derivative,
\begin{subequations}
\begin{align}
N_{V}^{\prime \mu} &= \Slash{D} V_\nu \Gamma^{\mu \nu}_{3/2} \gamma_5 q, \\
N_{A}^{\prime \mu} &= \Slash{D} A_\nu^i \Gamma^{\mu \nu}_{3/2}\tau^i q, \\
\Delta_{A}^{\prime \mu i} &= \Slash{D}  A_\nu^j \Gamma^{\mu \nu}_{3/2}P^{ij}_{3/2} q,
\end{align}%
\label{Mar0209eq1}%
\end{subequations}%
where $D_\mu$ denotes a covariant derivative. The mirror fields
$N_2^\mu$ and $\Delta_2^{\mu }$ are obtained by the same equations
as Eqs. (\ref{Sep2709eq1}) with substitution of the primed fields 
$(N_{V}^{\prime \mu}, N_{A}^{\prime \mu}, \Delta_{A}^{\prime \mu i})$
for the original fields $(N_{V}^{\mu}, N_{A}^{\mu}, \Delta_{A}^{\mu i})$. 
Although they would not be a unique possibility for the microscopic description of 
the mirror fields, Eqs.~(\ref{Mar0209eq1}) are enough for the present purpose to 
construct the chiral invariant Lagrangian. 

\section{Lagrangian}
\label{May2509sec2}
Now,  we proceed to the construction of the $SU(2)_R\times SU(2)_L$ Lagrangian. 
It is straightforward to show the chiral invariance of the kinetic terms:
${\cal L}_K= \bar{N}_{\nrm \mu} (i\ \Slash{\del}) N_\nrm^{\mu} + \bar{\Delta}_{\nrm \mu}^i (i\ \Slash{\del}) \Delta_\nrm ^{\mu i}, (\nrm=1, 2).$
In order to find interaction terms for higher-dimensional chiral multiplets, it 
is useful to employ the projection method.

\subsection{Mass terms and $\pi RR$ terms}

The vector and axial-vector diquarks belong to the chiral multiplet $(\half, \half)$, and 
$V_\mu^2+A_\mu^2$ is a chiral scalar. The Gell-Mann-Levy type interaction for the quark
$\bar{q} U_5 q$ is also a chiral scalar, where $U_5=\sigma +i\gamma_5 \inner{\tau}{\pi}$. 
Obviously, the following combination of these two terms is also a chiral scalar, 
\begin{eqnarray}
\bar{q} (V_\mu^2+A_\mu^2)U_5 q.
\label{eq:V2A2}
\end{eqnarray}
This term contains the direct products of the 
quark and diquark : $V^\mu q$ and $A^{\mu i} q$. They are decomposed into 
the irreducible parts as
\begin{subequations}
\begin{eqnarray}
& \left\{ \begin{array}{l}
V^\mu q =  \gamma_5 N_V^\mu + (J=\half \mbox{terms}), \\
A^{\mu i} q =
 \Delta_A^{\mu i} +\frac{1}{3} \tau^i N_A^\mu + (J=\half \mbox{terms}),
\end{array}\right. \\
& \left\{\begin{array}{l}
\bar{q} (V^\mu)^\dagger = -\bar{N}_{V }^\mu\gamma_5 + (J=\half \mbox{terms}), \\
\bar{q} (A^{\mu i})^\dagger =
\bar{\Delta}_{A}^{\mu i} +\frac{1}{3} \bar{N}_{A}^\mu \tau^i + (J=\half \mbox{terms}) ,
\end{array}\right.
\end{eqnarray}%
\label{eq:irrdcmp1}%
\end{subequations}%
Substituting Eqs.~(\ref{eq:irrdcmp1})  into the chiral invariant term~(\ref{eq:V2A2}),  we obtain
\begin{align}
{\cal L}_{MRR}^{(1)}=  & g_1\left( \bar{\Delta}_{1\mu}^i U_5 \Delta_1^{\mu i}
-\frac{3}{4}\bar{N}_{1\mu} U_5 N_1^\mu +\frac{1}{12} \bar{N}_{1\mu} \tau^i U_5 \tau^i N_1^\mu
+\frac{\sqrt{3}}{6}\left(\bar{N}_{1\mu} \tau^i U_5 \Delta_1^{\mu i}+\Hcterms \right) \right) \nonumber \\
& + (J=\half \mbox{terms}),
\label{eq:chiralint1}
\end{align}
where we omit $J=\half$ terms,  which contain the Gell-Mann-Levy type interaction 
with local nucleon operators $N_V =V_\mu \gamma^\mu q$ and 
$N_A =A_\mu^i \gamma^\mu \gamma_5 \tau^i q $. The transition terms between 
$J=\half$ and $\thalf$ fields vanish due to $\gamma_\mu \Delta_1^{\mu i}= \gamma_\mu N_1^\mu =0$.
The Lagrangian (\ref{eq:chiralint1}) describes several kinds of the interactions; 
the first three terms describe the diagonal interactions for $N_1^\mu$ and $\Delta_1^{\mu i}$ with 
$\sigma$ and $\pi$, and  the fourth term describes a transition between 
$N_1^\mu$ and $\Delta_1^{\mu i}$ with $\pi$, where a $\sigma N_1 \Delta_1$ coupling 
 vanishes due to $\tau^i \Delta_1^{\mu i}=0$. 

The diagonal interactions with $\sigma$ generate the masses of $N_1^\mu$ and 
$\Delta_1^{\mu i}$ in the presence of the SBCS $\sigma \to \bra \sigma \ket = f_\pi=92.4$ [MeV].
We obtain a mass relation $|m_{\Delta_1} | : | m_{N_1} |=2 : 1$. 
If we assign $N_1^\mu$ with $N(1520)$, which is the lowest lying state for $I(J) = \half(\thalf)$, 
its partner $\Delta_1^{\mu i}$ has the mass of $2\times 1520 \sim 3000$ MeV. 
We do not find a baryon suitable for this mass relation in the experimental data~\cite{Amsler:2008zzb}.

There are several directions to solve this mass problem: the inclusion of 
higher order terms in the Lagrangian and of higher-order diagrams, the extension 
of the chiral basis such as $(\thalf, 0)\oplus (0, \thalf)$ and  of the mirror 
assignment. It was shown ~\cite{Jido:1999hd} that the inclusion of the mirror 
assignment reasonably reproduces the masses and some properties of observed 
baryons. Using Eq.~(\ref{Mar0209eq2}), we find a chiral invariant interaction term
\begin{eqnarray}
{\cal L}_{MRR}^{(2)}=  g_2\left( \bar{\Delta}_{2\mu}^i U_5^\dagger \Delta_2^{\mu i}
-\frac{3}{4}\bar{N}_{2\mu} U_5^\dagger N_2^\mu +\frac{1}{12} \bar{N}_{2\mu} 
\tau^i U_5^\dagger \tau^i N_2^\mu
+\frac{\sqrt{3}}{6}\left(\bar{N}_{2\mu} \tau^i U_5^\dagger \Delta_2^{\mu i}
+{\rm H.c.}\right) \right),
\label{eq:eff2}
\end{eqnarray}
which is almost the same as Eq.~(\ref{eq:chiralint1}). The difference appears in the 
signs of the terms accompanying $\pi$ $(U_5\to U_5^\dagger)$, which is a feature of the 
mirror assignment~\cite{Jido:2001nt}.

Considering Eqs.~(\ref{Jul0608eq1}), (\ref{Sep2509eq1}) and (\ref{Mar0209eq2}), 
$\bar{\Delta}_{1R} \Delta_{2L}+ \bar{N}_{1R} N_{2L}$ is chiral invariant, which leads to
the following term,
\begin{eqnarray}
{\cal L}_{RR}=- m_0 \left(\bar{\Delta}_{1 \mu }^i \Delta_2^{\mu i}+
\bar{N}_{1\mu} N_2^\mu +{\rm H.c.}\right),
\label{eq:mirror}%
\end{eqnarray}%
which describes off-diagonal mass terms between $N_1^\mu$ and $N_2^\mu$ and 
between $\Delta_1^{\mu i}$ and $\Delta_2^{\mu i}$. The parameter $m_0$ 
describes a chiral scalar, so called mirror mass~\cite{Jido:2001nt}.

The mass terms included in ${\cal L}_{MRR}^{(1)}+{\cal L}_{MRR}^{(2)}+{\cal L}_{RR}$
are rewritten in the following matrix forms
\begin{eqnarray}
{\cal L}_M= - (\bar{\Delta}_{1\mu}^i, \bar{\Delta}_{2\mu}^i) 
\left( \begin{array}{cc}
-g_1 f_\pi & m_0 \\ m_0 & -g_2 f_\pi \end{array}\right) 
\left(\begin{array}{c} 
\Delta_1^{\mu i} \\ \Delta_2^{\mu i}
\end{array}\right)
-(\bar{N}_{1\mu}, \bar{N}_{2\mu})
\left(\begin{array}{cc} 
\half g_1 f_\pi & m_0 \\ m_0 & \half g_2 f_\pi 
\end{array}\right)
\left(\begin{array}{c}
N_1^\mu \\ N_2^\mu 
\end{array}\right).
\label{Mar0109eq1}
\end{eqnarray}
Because of the off-diagonal terms in these mass matrices, physical states and 
their masses are obtained through the diagonalization of the mass matrices.
Note that the mass eigen-values can take both positive and negative values. 
A state with a negative eigen-value can be transformed into a state with a 
positive mass, but has opposite parity to the original state. It is carried 
out by multiplying a state having negative mass by $\gamma_5$~\cite{Jido:2001nt}. 
In the present paper, we consider the case that two states form a pair of positive 
and negative parity states both in $\Delta$ and $N^*$ sectors.

For the $\Delta$ part in Eq.~(\ref{Mar0109eq1}), we obtain the mass eigen-values 
of two $\Delta$ states
\begin{align}
m_{\Delta^\pm}&= \half\left[ \sqrt{(g_1-g_2)^2 f_\pi^2+4m_0^2}\mp (g_1+g_2)f_\pi\right], 
\label{May2508eq1}
\end{align}
and the eigen-states
\begin{subequations}
\begin{align}
\Delta_+^{\mu i} &= \cos\theta_\Delta \Delta_1^{\mu i} +\sin\theta_\Delta \Delta_2^{\mu i},\\
\Delta_-^{\mu i} &= \gamma_5 (-\sin\theta_\Delta \Delta_1^{\mu i}+\cos\theta_\Delta \Delta_2^{\mu i}),
\label{Feb2109eq2}\\
\tan 2\theta_\Delta &=\frac{2m_0}{(g_2-g_1)f_\pi}.
\end{align}%
\label{Jul0508eq1}%
\end{subequations}%
Here we define  $\Delta_+^{\mu i}$ and $\Delta_-^{\mu i}$  as positive and negative 
parity states, respectively, where the indices $\pm$ denote the parity. Hence 
$\Delta_+^{\mu i}$ and $\Delta_-^{\mu i}$ are identified with $\Delta(P_{33})$ 
and $\Delta(D_{33})$, respectively. Note that $\gamma_5$ in Eq.~(\ref{Feb2109eq2}) 
appears due to the parity redefinition. Similarly, for $N^*$ part, we obtain the mass 
eigen-values
\begin{align}
m_{N^\pm} &= \half\left[ \sqrt{\frac{1}{4}(g_1-g_2)^2 f_\pi^2+4m_0^2} \pm \frac{(g_1+g_2)f_\pi}{2}\right],
\label{May2508eq2}
\end{align}
and the eigen-states 
\begin{subequations}
\begin{align}
N_+^{\mu}&=\cos\theta_N N_1^\mu +\sin\theta_N N_2^\mu,\\
N_-^{\mu}&=\gamma_5(-\sin\theta_N N_1^\mu+\cos\theta_N N_2^\mu),
\label{Feb2109eq3}\\
\tan 2\theta_N &=\frac{4m_0}{(g_1-g_2)f_\pi}.
\end{align}%
\label{Jul0508eq2}%
\end{subequations}%
$N_+^{\mu}$ and $N_-^{\mu}$ are identified with $N(D_{13})$ and $N(P_{13})$, 
respectively. Again, $\gamma_5$ in Eq.~(\ref{Feb2109eq3}) appears due to the 
parity redefinition. The four masses $m_{\Delta^{\pm}}$ and $m_{N^\pm}$ 
are given by the three parameters $g_1$, $g_2$ and $m_0$, which offers 
constraints on the four masses~\cite{Jido:1999hd},
\begin{subequations} 
\begin{align}
(m_{\Delta^+} + m_{\Delta^-}) & \geq (m_{N^+} + m_{N^-}), \\
 m_{\Delta^-} - m_{\Delta^+ } & =   2(m_{N^+} - m_{N^-}).
\end{align}%
\label{May1708eq1}%
\end{subequations}%
The inequality in the first line of Eq.~(\ref{May1708eq1}) is controlled 
by $m_0$. Thus, the mass splittings and average masses are determined 
by chiral symmetry and the mirror mass $m_0$.

It is worthwhile considering the correspondence between the basis states and 
the physical states. Obviously, the mixing angles vanish in the absence of 
the mirror mass; $\theta_N, \theta_\Delta \to 0$ for $m_0 \to 0$. In this limit, 
the naive and mirror sectors decouple, and  the physical states correspond to 
the basis states : $(\Delta_+^{\mu i}, N_+^\mu) \to (\Delta_1^{\mu i}, N_1^\mu)$ and 
$(\Delta_-^{\mu i}, N_-^\mu) \to (\Delta_2^{\mu i}, N_2^\mu)$. It should be 
noted that the decoupling of the two sectors does not violate chiral invariance. 
Contrarily, the two sectors are maximally mixed in the $m_0$ dominant case : 
$\theta_N, \theta_\Delta = \pi/4$.

The Lagrangians~(\ref{eq:chiralint1}) and (\ref{eq:eff2}) contain the one-pion 
interaction terms between the spin-$\thalf$ baryons ($\pi RR$) as well as the mass 
terms. Having the four spin-$\thalf$ baryons, there are ten coupling constants 
$g_{\pi RR}$; four diagonal and six off-diagonal terms. All the ten coupling 
constants are functions of  $g_1, g_2$ and $m_0$, which are determined by the masses.
It is straightforward to derive the $\pi RR$ coupling constants, $g_{\pi RR}$ 
from Eqs.~(\ref{eq:chiralint1}) and (\ref{eq:eff2}). For $\Delta$ part, we obtain
\begin{subequations}
\begin{align}
\Delta-\Delta & \left\{ \begin{array}{l}
 g_{\pi \Delta^+ \Delta^+}=-(g_1 \cos^2 \theta_\Delta -g_2 \sin^2 \theta_\Delta)  \\
 g_{\pi \Delta^- \Delta^-}=(g_1 \sin^2 \theta_\Delta -g_2 \cos^2 \theta_\Delta)    \\
 g_{\pi \Delta^+ \Delta^-}=(g_1+g_2)\cos\theta_\Delta \sin\theta_\Delta \\
\end{array}\right. 
\end{align}
which are defined by $\calL = -g_{\pi \Delta_P \Delta_{P^\prime}} \bar{\Delta}_{P \mu i} (i \gamma_5 \inner{\tau}{\pi}) \Gamma_5
\Delta_{P^\prime}^{\mu i}.$ Here $P$ and $P^\prime$ denote parity, i.e., $P, P^\prime = $ $+$ or $-$, 
and $\Gamma_5 = 1$ for $P=P^\prime$ and $\gamma_5$ for $P\neq P^\prime$. 
For $N^*$ part, we obtain
\begin{align}
N^*-N^* & \left\{ \begin{array}{l}
 g_{\pi N^+ N^+}= \frac{5}{6} (g_1 \cos^2 \theta_N -g_2 \sin^2 \theta_N) \\
 g_{\pi N^- N^-}= -\frac{5}{6}(g_1 \sin^2 \theta_N -g_2 \cos^2 \theta_N) \\
 g_{\pi N^+ N^-}= -\frac{5}{6}(g_1+g_2)\cos\theta_N \sin\theta_N \\
\end{array}\right. 
\end{align}
which are defined by $\calL = -g_{\pi N_P N_{P^\prime}} \bar{N}_{P \mu} (i \gamma_5 \inner{\tau}{\pi}) \Gamma_5 N_{P^\prime}^{\mu}$.  
For $N^*$-$\Delta$ transition terms, 
\begin{align} 
N^*-\Delta & \left\{ \begin{array}{l}
 g_{\pi N^+ \Delta^+}= -\frac{\sqrt{3}}{3} (g_1 \cos \theta_\Delta \cos \theta_N -g_2 \sin \theta_\Delta \sin \theta_N ) \\
 g_{\pi N^+ \Delta^-}= \frac{\sqrt{3}}{3} (g_2 \cos \theta_\Delta \sin \theta_N +g_1 \cos \theta_N \sin \theta_\Delta )  \\
 g_{\pi N^- \Delta^+}= - \frac{\sqrt{3}}{3} (g_1 \cos \theta_\Delta \sin \theta_N +g_2 \cos \theta_N \sin \theta_\Delta ) \\
 g_{\pi N^- \Delta^-}= \frac{\sqrt{3}}{3} (g_1 \sin \theta_\Delta \sin \theta_N -g_2 \cos \theta_N \cos \theta_\Delta ) \\
\end{array}\right.
\end{align}
which are defined by $\calL = -g_{\pi N_P \Delta_{P^\prime}} \bar{N}_{P \mu} (i \gamma_5 \Gamma_5 ) \pi^i \Delta_{P^\prime}^{\mu i}$.  
\end{subequations}
In order to understand the features of $g_{\pi RR}$, it is 
useful to consider the  axial-charges, which are obtained by the Noether theorem 
\begin{align}
\Delta-\Delta & \left\{ \begin{array}{l}
g_A^{\Delta^\pm \Delta^\pm} = \pm \cos 2\theta_\Delta, \\
g_A^{\Delta^+ \Delta^-} = -\sin 2\theta_\Delta, 
\end{array}\right. \nn \\
N^*-N^* & \left\{ \begin{array}{l}
g_{A}^{N^{*\pm} N^{*\pm}} =\pm \frac{5}{3} \cos 2\theta_N, \\
g_A^{N^{*+} N^{*-}} =- \frac{5}{3} \sin 2\theta_N, 
\end{array}\right. \nn \\ 
N^*-\Delta & \left\{ \begin{array}{l}
g_A^{N^{*\pm} \Delta^\pm} =\pm\frac{4}{\sqrt{3}} \cos (\theta_N+\theta_\Delta), \\
g_A^{N^{*\pm} \Delta^\mp} =\pm\frac{4}{\sqrt{3}} \sin (\theta_N+\theta_\Delta).
\end{array}\right.
\label{Nov0209eq1}%
\end{align}%
In the limit $\theta_{N,\Delta}\to 0$ ($m_0\to 0$),  the absolute values of the parity-non-changing 
interactions reach the maximum values: 
$|g_A^{\Delta^\pm \Delta^\pm} |\to 1$, $|g_A^{N^{*\pm} N^{*\pm}}| \to \frac{5}{3}$ and 
$|g_A^{N^{*\pm} \Delta^\pm} |\to \frac{4}{\sqrt{3}}$, while the parity-changing terms vanish 
$g_A^{\Delta^+ \Delta^-}=g_A^{N^+ N^-} =g_A^{N^\pm \Delta^\mp}=0$.
 The mixing angles larger, as $m_0$ becomes larger.
Since the naive and mirror sectors have the opposite axial-charges, 
the mixing of the two sectors suppresses the parity-non-changing interactions
and enhance the parity-changing interactions.
In the $m_0$-dominance, the parity-non-changing interactions vanish
$g_A^{\Delta^\pm \Delta^\pm} = g_A^{N^{*\pm} N^{*\pm}} = g_A^{N^{*\pm} \Delta^\pm} \to 0$, 
while the parity-changing terms reach the maximum values $|g_A^{\Delta^+ \Delta^-}|= 1$, 
$|g_A^{N^+ N^-}| = \frac{5}{3}$ and $|g_A^{N^\pm \Delta^\mp}|=\frac{4}{\sqrt{3}}$.
Of course,  $g_{\pi RR}$ have the same features as the axial-charges due to 
 the Goldberger-Treiman (GT) relations:  
\begin{align}
\Delta - \Delta & \left\{
\begin{array}{l}
f_\pi g_{\pi \Delta^+ \Delta^+}  =  \cos 2\theta_\Delta m_{\Delta^+},\\
f_\pi g_{\pi \Delta^- \Delta^-}  =  -\cos 2\theta_\Delta m_{\Delta^-},\\
f_\pi g_{\pi \Delta^+ \Delta^-}  =  -\frac12 \sin 2\theta_\Delta (m_{\Delta^+}-m_{\Delta^-}),
\end{array}\right. \nn \\
N^* - N^* & \left\{  
\begin{array}{l}
f_\pi g_{\pi N^+ N^+}  = \frac{5}{3} \cos 2\theta_N m_{N^+},\\
f_\pi g_{\pi N^- N^-}  = -\frac{5}{3} \cos 2\theta_N m_{N^-},\\
f_\pi g_{\pi N^+ N^-}  = -\frac{5}{6} \sin 2\theta_N (m_{N^+}-m_{N^-}),
\end{array}\right. \nn \\
N^*-\Delta &
\left\{\begin{array}{l}
f_\pi g_{\pi N^+ \Delta^+}  =  \frac{2}{\sqrt{3}} \cos (\theta_N+\theta_\Delta) (m_{N^+}+m_{\Delta^+}) ,\\
f_\pi g_{\pi N^+ \Delta^-}  = -\frac{2}{\sqrt{3}} \sin (\theta_N+\theta_\Delta) (m_{N^+}-m_{\Delta^-}) ,\\
f_\pi g_{\pi N^- \Delta^+}  = -\frac{2}{\sqrt{3}} \sin (\theta_N+\theta_\Delta) (m_{N^-}-m_{\Delta^+}) ,\\
f_\pi g_{\pi N^- \Delta^-}  = -\frac{2}{\sqrt{3}} \cos (\theta_N+\theta_\Delta) (m_{N^-}+m_{\Delta^-}).
\end{array}\right.
\end{align}

\subsection{Interaction with the nucleon}

Next, we construct the interactions between the nucleon $(N)$ and the chiral quartet. As we have 
discussed in the introduction, we assume that the nucleon belongs to $(\half, 0)\oplus (0, \half)$. 
With the nucleon's chiral multiplet,  we can classify the products of the chiral properties of $N\otimes \Delta$:
\begin{align}
N\otimes \Delta & = \left[\left(\half, 1\right)\oplus \left(1, \half\right )\right]\otimes \left[\left(\half,0\right)\oplus \left(0, \half\right)\right] \nn \\ 
&= \left\{ \begin{array}{lcl}   (1, 0)\oplus (0, 1)  & \mbox{for}  & (N_1^\mu, \Delta_1^{\mu i}), \\
  \left(\half, \half\right)  & \mbox{for} & (N_2^\mu, \Delta_2^{\mu i}),
\end{array}\right.
\label{Feb0210eq1}
\end{align}
where we omit four-meson terms $(1,1)$ and $\left[ \left(\thalf,\half\right)\oplus \left(\half, \thalf\right)\right]$.
In the derivation of Eq.~(\ref{Feb0210eq1}), it is important to take into account the chirality conservation.
This classification implies that  chiral invariant interactions between $N$ and 
$(N_1^\mu, \Delta_1^{\mu i})$ accompany two-meson fields, 
while those between $N$ and  $(N_2^\mu, \Delta_2^{\mu i})$ 
accompany one-meson fields.

We find two chiral scalars $\sigma V_\mu + i\inner{\pi}{A}_\mu$ and
$\bar{N} U_5 q $. Multiplying them,  we find two chiral invariant terms;
$(-i) \bar{N}  U_5 [ ( \del^\mu\sigma) V_\mu +i  \inner{(\del^\mu\pi)}{A_\mu}]q$ , 
$(-i) \bar{N} ( \del^\mu U_5) (\sigma V_\mu + i\inner{\pi}{A_\mu})q$.  
Using Eqs.~(\ref{eq:irrdcmp1}), we obtain the chiral invariant interaction terms
between $N$ and $(N^\mu_1,\; \Delta^{\mu i}_1)$
\begin{subequations}
\begin{eqnarray}
{\cal L}_{M N R}^{(1)}&=& \frac{g_3}{\Lambda^2} \left[\bar{N} O_{1\mu}^i \Delta_1^{\mu i}
+\bar{N} O_{2 \mu} N_1^\mu\right]+ \Hcterms, 
 \label{Feb2509eq1}\\
{\cal L}_{M N R}^{(2)} &=& \frac{g_4}{\Lambda^2} \left[\bar{N} O_{3\mu }^i \Delta_1^{\mu i}
+\bar{N} O_{4\mu} N_1^\mu\right]+ \Hcterms, 
\end{eqnarray}%
\label{Nov0209eq2}
where the dimensional parameter $\Lambda$ [mass] is introduced to keep 
the coupling constants $g_3$ and $g_4$ dimensionless.  
We also introduce shorthand notations $O_\nrm\; (\nrm = 1, \cdots 4)$ for mesonic operators
\begin{align}%
O_{1}^{\mu i} &=U_5 (\del^\mu \pi^i), \\
O_{2}^\mu  & = - \frac{\sqrt{3}}{2} U_5 \left( ( \del^\mu \sigma)\gamma_5 + \frac{1}{3}  (i \del^\mu \inner{\pi}{\tau})\right), \\
O_{3}^{\mu i} &= (\del^\mu U_5) ( \pi^i), \\
O_{4}^\mu &= - \frac{\sqrt{3}}{2}(i \del^\mu U_5) \left(\sigma\gamma_5 +\frac{1}{3}i \inner{\pi}{\tau}\right).
\end{align}%
\end{subequations}%
One may think it possible to construct similar interaction terms for the mirror fields 
by the replacement Eq.~(\ref{Mar0209eq2}). However, such terms are forbidden by chirality 
conservation, as is shown in Eq.~(\ref{Feb0210eq1})~\footnote{It can be shown explicitly.
For example, the first term in Eq.~(\ref{Feb2509eq1})  is rewritten in terms of left- 
and right-handed parts of the fields as
$\bar{N} U_5 ( \del_\mu \pi ^i)\Delta_1^{\mu i} = \bar{N}_L U_5 ( \del_\mu \pi ^i) \Delta_{1R} + (l\leftrightarrow r).$
Replacing $\Delta_{1R} \to \Delta_{2L}$,  $\bar{N}_L U_5 (\del_\mu \pi ^i) \Delta_{1R} \to \bar{N}_L U_5(\del_\mu \pi ^i) \Delta_{2L}$, 
which vanishes due to $P_L P_R =0, (P_{R, L} = (1\pm \gamma_5)/2)$.}.  
The mirror fields have one-meson interactions with the nucleon.
It can be constructed by using the chiral invariant operators $(-i)(\sigma V_\mu + i \inner{\pi}{A}_\mu)$ and $\bar{N} \Slash{D} q$.
We obtain
\begin{subequations}
\begin{eqnarray}
{\cal L}_{M N R}^{(3)}&=& \frac{g_5}{\Lambda}\left[\bar{N} O_{5\mu}^i \Delta_{2}^{\mu i}
+ \bar{N} O_{6\mu} N_2^\mu
\right], 
\end{eqnarray}
where $O_5$ and $O_6$ are also mesonic operators,
\begin{align}
O_{5}^{\mu i}  &= (\del^\mu \pi^i), \\
O_{6}^ {\mu}  &= - \frac{\sqrt{3}}{2}(i\del^\mu)(\sigma\gamma_5+\frac{1}{3} i \inner{\tau}{\pi}).
\end{align}%
\end{subequations}%
In the mass basis, ${\cal L}_{MNR}={\cal L}_{MNR}^{(1)}+{\cal L}_{MNR}^{(2)}+{\cal L}_{MNR}^{(3)} $ is rewritten as 
\begin{align}
{\cal L}_{MNR} & =  
\bar{N} \left[ (O_{1\mu}^i + O_{3\mu}^i) \cos \theta_{\Delta} + O_{5\mu}^i \sin \theta_{\Delta} \right] \Delta_+^{\mu i} 
+ \bar{N} \left[- (O_{1\mu}^i + O_{3\mu}^i) \sin \theta_{\Delta} + O_{5\mu}^i 
\cos\theta_{\Delta} \right] \gamma_5 \Delta_-^{\mu i}\nn \\
&+\bar{N} \left[ (O_{2\mu} + O_{4\mu}) \cos \theta_N + O_{6\mu} \sin \theta_N \right] N_+^{\mu i}
+ \bar{N} \left[ -(O_{2\mu} + O_{4\mu}) \sin \theta_N + O_{6\mu} \cos \theta_N \right]\gamma_5 N_-^{\mu i}, 
\label{Sep1909eq1}
\end{align}
which contains several kinds of the interaction terms, $\pi N R$, $\pi\pi N R$, $\sigma N R$ and 
$\sigma\sigma NR$. Among them,  we consider $\pi NR$ and $\pi \pi NR$ interaction terms 
in order for the comparison with experiments.
The $\pi N$ interactions of the chiral quartet are given by
\begin{subequations}
\begin{align}
{\cal L}_{\pi N R}&= \frac{g_{\pi N\Delta^+}}{\Lambda} \bar{N} (\del_\mu \pi^i)\Delta^{+\mu i}+ 
\frac{g_{\pi N\Delta^-}}{\Lambda} \bar{N} (\del_\mu \pi^i)\gamma_5 \Delta^{-\mu i} \nn\\
&+\frac{g_{\pi N N^{*-}}}{\Lambda} \bar{N} (\del_\mu \inner{\pi}{\tau})\gamma_5 N^{-\mu}
+\frac{g_{\pi N N^{*+}}}{\Lambda} \bar{N} (\del_\mu \inner{\pi}{\tau} ) N^{+\mu},
\label{Mar0209eq3}
\end{align}
where the coupling constants $g_{\pi N N^{* \pm} }$ and $g_{\pi N\Delta^\pm}$ are given by 
\begin{align}
g_{\pi N\Delta^+} & = \frac{1}{\Lambda} (g_5 \Lambda \sin \theta_\Delta 
+ g_3 f_\pi \cos \theta_\Delta ),\\
g_{\pi N\Delta^-} & = \frac{1}{\Lambda} (g_5 \Lambda \cos \theta_\Delta 
- g_3 f_\pi \sin \theta_\Delta ),\\
g_{\pi N N^{* +} } & = \frac{\sqrt{3}}{6\Lambda} (g_5 \Lambda 
\sin \theta_N  + (g_3 +3 g_4 )f_\pi \cos \theta_N),\\
g_{\pi N N^{* -} } & = \frac{\sqrt{3}}{6\Lambda} (g_5 \Lambda 
\cos \theta_N  - (g_3 +3 g_4 )f_\pi \sin \theta_N).
\end{align}%
\label{Feb2509eq2}%
\end{subequations}%
Four $g_{\pi NR}$ are expressed  in terms of three parameters $g_3, g_4$ and $g_5$, which leads to
one identity 
\begin{align}
(\sin \theta_\Delta g_{\pi N \Delta^+} + \cos \theta_\Delta g_{\pi N \Delta^-} ) = 2 \sqrt{3} ( \sin \theta_N g_{\pi N N^{* +}} + \cos \theta_N g_{\pi N N^{* -}} ).
\label{Feb2509eq3}
\end{align}
Here it must be noted that the derivation of the $\pi N$ interactions is based on the assumption of the 
nucleon's chiral multiplet. If the nucleon together with the negative parity resonance 
group into $(\half,0)\oplus (0, \half)$ with the mirror assignment, 
we can include three additional interactions, which spoils the constraint Eq.~(\ref{Feb2509eq3}).
Another possibility is that the nucleon contains $(1,\half)\oplus (\half,1)$ as well as 
$(\half,0)\oplus (0,\half)$. In this case, we can include one additional interaction that have 
similar form to Eq.~(\ref{eq:chiralint1}). With the new term, Eq.~(\ref{Feb2509eq3}) becomes 
loose constraint and gives the ordering of the coupling constants. So, Eq.~(\ref{Feb2509eq3}) 
is one of the most strict constraint. The point is that it is possible to improve this result 
without changing the masses and $\pi RR$ interactions of the quartet.

We obtain two-pion interaction terms
\begin{align}
 {\cal L}_{\pi \pi N \Delta} & =  \frac{g_{\pi\pi N\Delta_+ }^{(v)}}{\Lambda} \bar{N} ( \epsilon^{abc}\pi^a \pi_{,\mu}^b \gamma_5 ) \Delta_+^{\mu c}
\; + \;  \frac{g_{\pi \pi N \Delta^+ }^{(t)}}{\Lambda} \bar{N} (\pi^a \pi_{,\mu}^b + \pi_{,\mu}^a \pi^b) (i \gamma_5 \tau^a) \Delta_+^{\mu b} 
  \nn \\
& +  \frac{g_{\pi\pi N \Delta_- }^{(v)} }{\Lambda} \bar{N} ( \epsilon^{abc}\pi^a \pi_{,\mu}^b ) \Delta_-^{\mu c}
  \; + \;  \frac{g_{\pi \pi N \Delta_- }^{(t)}}{\Lambda} \bar{N} (\pi^a \pi_{,\mu}^b + \pi_{,\mu}^a \pi^b) (i \tau^a) \Delta_-^{\mu b} 
\label{Feb0310eq1} \\
{\cal L}_{\pi \pi N N^*} &=  \frac{g_{\pi\pi N N_+^*}^{(s)}}{\Lambda} \bar{N} (i \gamma_5 \inner{\pi}{\pi_{,\mu}} ) N_+^\mu 
    \; + \;  \frac{g_{\pi \pi N N_+^*}^{(v)} }{\Lambda} \bar{N} ( \epsilon^{abc} \pi^a \pi^b_{,\mu} \tau^c) \gamma_5 N_+^\mu \nn \\
   & + \frac{g_{\pi\pi N N_-^*}^{(s)} }{\Lambda}  \bar{N} (i \inner{\pi}{\pi_{,\mu}} ) N_-^\mu 
    \; + \;  \frac{g_{\pi \pi N N_-^*}^{(v)}}{\Lambda} \bar{N} ( \epsilon^{abc} \pi^a \pi^b_{,\mu} \tau^c)  N_-^\mu, 
    \label{Feb0310eq2}
\end{align}
with 
\begin{align}
\Delta\mbox{-sector} \left\{ \begin{array}{ll}
g_{\pi \pi N \Delta_+ }^{(v)} &= \frac{\cos \theta_\Delta}{2 \Lambda} (g_3 - g_4) , \\
g_{\pi \pi N \Delta^+ }^{(t)} &= \frac{\cos \theta_\Delta}{ 2\Lambda} (g_3 + g_4), \\
g_{\pi \pi N \Delta_- }^{(v)} &= - \frac{\sin \theta_\Delta}{2 \Lambda} (g_3 - g_4), \\
g_{\pi \pi N \Delta_- }^{(t)} &= - \frac{\sin \theta_\Delta}{2 \Lambda} (g_3 + g_4), 
\end{array}\right. 
\;\; \; N^* \mbox{-sector} \left\{ \begin{array}{ll}
      g_{\pi\pi N N_+^*}^{(s)} &= +\frac{\sqrt{3} \cos \theta_N}{6 \Lambda} (g_3 + g_4) , \\
      g_{\pi \pi N N_+^*}^{(v)}&= -\frac{\sqrt{3} \cos \theta_N}{6 \Lambda} (g_3 - g_4) ,\\
      g_{\pi\pi N N_-^*}^{(s)} &= - \frac{\sqrt{3} \sin \theta_N}{6 \Lambda} (g_3 + g_4),\\
      g_{\pi \pi N N_-^*}^{(v)}&= \frac{\sqrt{3} \sin \theta_N}{6 \Lambda} (g_3 - g_4), 
\end{array}\right.
\label{Feb0310eq3}%
\end{align}%
where they are classified into three types: the symmetric ($\inner{\pi}{\pi_{,\mu}}$), 
anti-symmetric $(i \epsilon^{abc} \pi^a \pi^b_{,\mu})$ and  symmetric type 
$(\pi^a\pi^b_{,\mu} + \pi^a_{,\mu} \pi^b)$. They corresponds to an isoscalar ($\inner{\pi}{\pi_{,\mu}}$), 
isovector  $(i \epsilon^{abc} \pi^a \pi^b_{,\mu})$ and isotensor $(\pi^a\pi^b_{,\mu} + \pi^a_{,\mu} \pi^b)$.
Since the two-pion coupling constants $g_{\pi\pi N R}$ contain only $g_3$ and $g_4$, 
their strengths are determined by the $\pi N$ coupling constants through
$g_3 = (\Lambda/f_\pi) ( (g_{\pi N \Delta^+} - g_{\pi N \Delta^-})/ (\cos \theta_\Delta + \sin \theta_\Delta))$
and $g_4 = (2\Lambda/\sqrt{3}f_\pi ) ( (g_{\pi N N^*_+} - g_{\pi N N^*_-})/( \cos \theta_N + \sin \theta_N))$.
Furthermore,  $g_{\pi\pi NR}$ are proportional to either $(g_3+g_4)$ or $(g_3-g_4)$,
which provides a selection rule; either $\pi\pi$ isoscalar or isovector interaction is suppressed each for $N^*_\pm$,
and either the isovector or isotensor interaction is suppressed each for $\Delta_\pm$. 

Using the $SU(2)_R\times SU(2)_L$ Lagrangian, we have derived several constraints on 
the properties of the chiral quartet. We concentrate on the construction of the 
lowest-order terms and the derivation of the chiral constraints at tree level.
In general, it is possible to insert chiral invariant operators such as $(\sigma^2+\pi^2)^n$ 
into the chiral Lagrangians we derived. However, those terms does not change the 
above constraints and can be absorbed into the parameters. Regarding the 
$\pi RR$ interactions, it is possible to include additional interaction term with 
a derivative~\cite{Jido:2001nt}. The constraint for the $\pi NR$ interactions rely
 on the assumption of the saturation of $(\half, 0)\oplus (0, \half)$ in the nucleon. 
The inclusion of $(1,\half)\oplus (\half,1)$ component in the nucleon causes 
 one additional chiral invariant $\pi N$ interaction term similar to Eq.~(\ref{eq:chiralint1}). 
In this case,  four $g_{\pi NR}$ are given by four parameters.
It must be noted that the inclusion of $(1,\half)\oplus (\half,1)$ for the nucleon
does not affect the multiplet nature of the quartet.

\section{Results}
\label{May2509sec3}

\begin{table}[htbp]
\begin{center}
\begin{minipage}{10cm}
\caption{Observed states listed in PDG~\protect\cite{Amsler:2008zzb} 
corresponding to the quantum numbers of the members of the quartet.
The number of the stars denotes PDG-ratings of the states.}
\label{May1408tab1}
\end{minipage}
\begin{tabular}{cl}
\hline \hline
  $L_{2I 2J}$ & Observed states\\
 \hline
 $P_{33}$ & $\Delta(1232)^{****}$, $\Delta(1600)^{***}$, $\Delta(1920)^{***}$ \\
 $D_{33}$ & $\Delta(1700)^{***}$, $\Delta(1940)^{**}$ \\
 $D_{13}$ & $N(1520)^{****}$, $N(1700)^{***}$, $N(2080)^{**}$ \\
 $P_{13}$ & $N(1720)^{****}$, $N(1900)^{*}$ \\
\hline\hline
\end{tabular}
\end{center}
\end{table}

\begin{table}[htbp]
\begin{center}
\begin{minipage}{12cm}
\caption{Data for masses, $\pi N$ decay widths and 
$\pi N$ coupling constants of the observed states used in 
the cases (1) and (2). The data are taken from PDG~\protect\cite{Amsler:2008zzb}. 
The values in the bracket for $m_R^{\rm(exp)}$ are central values of 
the observed masses, while those for $\Gamma_{\pi N}^{\rm (exp)}$ 
are the average values between minimum and maximum values. 
The definition of $g_{\pi N}^{\rm (exp)}$ is given in the main text. 
For $\Delta(1940)$ in the case (2), we use the data in Ref.\protect~\cite{Horn:2008qv}.}
\label{May1910tab1}
\vspace{0.2cm}
\end{minipage}
\begin{tabular}{ccccc}
\hline \hline
 States $R$  & $m_R^{\rm(exp)}$ [MeV] & $\Gamma_{\pi N}^{\rm (exp)}$ [MeV] & 
$g_{\pi N}^{\rm (exp)}/ \Lambda$ [GeV$^{-1}]$ \\
 \hline
 $\Delta(1232) [P_{33}]$ & 1231-1233 (1232) & 116-120   (118)  & 15.7-16.0 (15.8) \\
 $\Delta(1600) [P_{33}]$ & 1550-1700 (1600) & 25.0-113  (68.8) & 2.37-5.04 (3.70) \\
 $\Delta(1700) [D_{33}]$ & 1670-1750 (1700) & 20.0-80.0 (50.0) & 6.34-12.7 (9.51) \\
 $\Delta(1940) [D_{33}]$ & 1950-2030 (1990) & 17.0-62.4 (39.7) & 3.23-6.20 (4.72) \\
 $N(1520) [D_{13}]$      & 1515-1525 (1520) & 55.0-81.3 (68.1) & 7.64-9.30 (8.46) \\
 $N(1720) [P_{13}]$      & 1700-1750 (1720) & 15.0-60.0 (37.5) & 1.72-3.44 (2.58) \\
 \hline  \hline
\end{tabular}
\end{center}
\end{table}

\begin{table}[htbp]
\begin{center}
\begin{minipage}{12cm}
\caption{
Data for masses, $\pi N$ decay widths and 
$\pi N$ coupling constants of the observed states used in 
the cases (3-1) and (3-2). See also the caption of Table~\ref{May1910tab1}.}
\label{May2510tab1}
\vspace{0.2cm}
\end{minipage}
\begin{tabular}{ccccl}
\hline \hline
 \multicolumn{5}{c}{Case (3-1)} \\
 States $R$  & $m_R^{\rm (exp)}$ [MeV] & $\Gamma_{\pi N}^{\rm (exp)}$ [MeV] & $g_{\pi N}^{\rm (exp)}/ \Lambda $ [GeV$^{-1}]$
 & Reference\\
 \hline
 $\Delta(1920) [P_{33}]$ & 1900-1970 (1920) & 7.50-60.0 (33.8) & 0.825-2.33(1.58) & PDG average~\cite{Amsler:2008zzb}     \\
 $\Delta(1940) [D_{33}]$ & 1950-2030 (1990) & 17.0-62.4 (39.7) & 3.23-6.20(4.72)  & Horn et. al.~\cite{Horn:2008qv} \\
 $N(2080) [D_{13}]$      & 1945-1947 (1946) & 85.2-121  (103)  & 4.63-5.23(5.08)  & Penner et. al.~\cite{Penner:2002ma}   \\
 $N(1900) [P_{13}]$      & 1855-1975 (1915) & 2.80-19.8 (11.3) & 0.574-1.53(1.05) & Nikonov et. al.~\cite{Nikonov:2007br} \\
\hline  \hline 
 \multicolumn{5}{c}{Case (3-2)} \\
 States $R$  & $m_R^{\rm (exp)}$ [MeV] & $ \Gamma_{\pi N}^{\rm (exp)}$ [MeV] & $g_{\pi N}^{\rm (exp)} / \Lambda$ [GeV$^{-1}]$
 & Reference\\
 \hline
 $\Delta(1920) [P_{33}]$ & 1900-1970 (1920) & 7.50-60.0  (33.8)  & 0.825-2.33 (1.58) & PDG average~\cite{Amsler:2008zzb}  \\
 $\Delta(1940) [D_{33}]$ & 1947-2167 (2057) & 8.40-234   (121)   & 2.04-10.8  (6.40) & Manley et. al.~\cite{Manley:1992yb} \\
 $N(2080) [D_{13}]$      & 1749-1859 (1804) & 53.0-165   (109)   & 4.45-7.84  (6.15) & Manley et. al.~\cite{Manley:1992yb} \\
 $N(1900) [P_{13}]$      & 1855-1975 (1915) & 2.8.0-19.8 (11.3)  & 0.574-1.53 (1.05) & Nikonov et. al.~\cite{Nikonov:2007br}\\
\hline  \hline
\end{tabular}
\end{center}
\end{table}
In this section, we proceed to numerical discussions and look for a set of baryons suitable 
for the QS. Possible candidates for the members of the quartet are shown in Table~\ref{May1408tab1}. 
There are six parameters in our model: $m_0, g_1$, $g_2$, $g_3, g_4$ and $g_5$. 
The dimensional parameter $\Lambda$ does not play any role in the present study, 
then we do not need to determine it. Since the masses $m_{\Delta_\pm}$ and 
$m_{N^*_\pm}$ are the functions of $m_0, g_1$, and $g_2$, we can determine 
them by minimizing $\chi^2_{\rm mass} = \sum_R (m_R-m_R^{{\rm (exp)}})^2/ (\delta m_R^{\rm (exp)})^2, 
( R= \Delta_\pm$ and $N^*_\pm)$. Here $m_R^{\rm (exp)}$ and $\delta m_R^{\rm (exp)}$ 
are the central values and errors of the observed masses, which are shown in Table~\ref{May1910tab1} 
and \ref{May2510tab1}. Considering the sates listed in Table~\ref{May1408tab1}, 
there are 36 possible assignments.  Among them, we discuss four cases  
{\bf [Case (1)] $(\Delta(1232)$, $\Delta(1700)$, $N(1520)$, $N(1720))$, 
[Case (2)] $(\Delta(1600)$, $\Delta(1940)$, $N(1520)$, $N(1720))$}, 
{\bf [Case (3-1)]}{ and {\bf [(3-2)]} $(\Delta(1920)$, $\Delta(1940)$, $N(2080)$, $N(1900)$). 
Although the case (1) was studied in Ref.~\cite{Jido:1999hd,Nagata:2008xf}, 
we reanalyze this case with the use of the different method for the determination of 
the parameters. As we will show, the case (2) agrees with the mass pattern of the QS with the 
smallest $\chi^2_{\rm mass}$. We also discuss ($\Delta(1920)$, $\Delta(1940)$, $N(2080)$, $N(1900)$). 
Because of a variety in the data, we consider two cases, (3-1) and (3-2), for this assignment, 
using two different data sets shown in Table~\ref{May2510tab1}. 
There are three other assignments that reproduce the 
masses with $\chi^2_{\rm mass}$ less than one: ($\Delta(1600)$, $\Delta(1700)$, $N(1700)$, $N(1720)$), 
($\Delta(1600)$, $\Delta(1940)$, $N(1700)$, $N(1900)$), 
($\Delta(1920)$, $\Delta(1940)$, $N(1700)$, $N(1720)$). 
We concentrate on the above four cases in the present work.
Instead of discussing all of them, we discuss the general behaviors of the QS later. 
Results for the masses are shown in Table~\ref{Jun2708tab1}. 
\begin{table}[htbp]
\begin{center}
\begin{minipage}{12cm}
\caption{Result for the masses and parameters. 
For the experimental data, see Table~\ref{May1910tab1} and \ref{May2510tab1}.}
\label{Jun2708tab1}
\vspace{0.2cm}
\end{minipage}
\begin{tabular}{ccccc}
\hline \hline
  & \multicolumn{4}{c}{Masses  [MeV]  [Assigned states] } \\
         State  &  Case (1) &  Case (2) & Case (3-1) & Case (3-2)\\
\hline
 $\Delta^{+}$ [$P_{33}$] & 1233 [$\Delta(1232)$]  &  1594 [$\Delta(1600)$] & 1935 [$\Delta(1920)$] & 1917 [$\Delta(1920)$] \\
 $\Delta^{-}$ [$D_{33}$] & 2190 [$\Delta(1700)$]  &  1992 [$\Delta(1940)$] & 1980 [$\Delta(1940)$] & 2083 [$\Delta(1940)$] \\
 $N^{-}$  [$D_{13}$]     & 1473 [$N(1520)$]       &  1520 [$N(1520)$]      & 1946 [$N(2080)$]      & 1817 [$N(2080)$] \\
 $N^{+}$  [$P_{13}$]     & 1951 [$N(1720)$]       &  1719 [$N(1720)$]      & 1969 [$N(1900)$]      & 1899 [$N(1900)$] \\
  \hline 
 $\chi^2_{\rm mass}$     &     68      &  0.0025     &  0.26      &  0.045     \\
\hline  \hline 
   &           \multicolumn{4}{c}{Parameters and angles} \\
 State  &  Case (1) &  Case (2)  & Case (3-1) & Case (3-2) \\
\hline
$g_1$                      &   5.2    &   12   & 0.25 &  10  \\
$g_2$                      &   5.2    &  -7.5  & 0.25 & -8.3 \\
$m_0$ [MeV]                &  1712    &  1557  & 1957 & 1809 \\
$\theta_N$ [degree]        &   45     &    37  &   45 &  38  \\
$\theta_\Delta$ [degree]   &   45     &    60  &   45 &  58  \\
  \hline \hline
\end{tabular}
\end{center}
\end{table}
For the case (1), the present result differs from the previous study~\cite{Nagata:2008xf}, 
which is due to the difference of the method to determine the mass parameters. 
In Ref.~\cite{Nagata:2008xf}, we adopted the minimization of a standard deviation 
$\sigma^2 = \sum_R (m_R -m_R^{{\rm (exp)}})^2$, while we employ 
$\chi^2$-minimum method in the present work. These two methods differ in how 
$\Delta(1232)$ are included in the fitting procedure, because the error of the 
observed $\Delta(1232)$'s mass is much smaller than those of the other three states. 
We found $\chi^2_{\rm mass}$ amounts to 60, which is significantly large. 
It is favorable for the QS that the masses of the $\Delta_\pm$ are larger than those of 
$N^*_\pm$, as shown in Eqs.~(\ref{May1708eq1}). The mass of $\Delta(1232)$ is much smaller 
compared with other spin-$\thalf$ baryons. This causes the significantly large discrepancy. 
We also found that $\chi^2_{\rm mass}$ becomes larger if assignments include 
$\Delta(1232)$ as a member of the quartet, which implies that the mass of 
$\Delta(1232)$ is too small for the QS. 

The cases (2), (3-1) and (3-2) are new in this work. The case (2) is the best assignment 
for the quartet with $\chi^2_{\rm mass} =0.0025$, which is the smallest value among $\chi^2_{\rm mass}$ for 
36 possible assignments. For $\Delta(1940)$ in this case, we use the data by  Horn et. al.~\cite{Horn:2008qv}.
We confirmed that the result for (2) is insensitive to the choice of the data for $\Delta(1940)$. 
The cases (3-1) and (3-2) also reproduce the masses of the quartet with $\chi^2_{\rm mass}$ = 0.26 and 
0.045, respectively. 

\begin{table}[htbp]
\begin{center}
\begin{minipage}{12cm}
\caption{The one-pion coupling constants between the the members 
of the quartet, $g_{\pi RR}$. The values of the parameters are 
shown in Table~\ref{Jun2708tab1}.}
\label{May2708tab1}
\end{minipage}
\begin{tabular}{lcccc}
\hline \hline
 $g_{\pi RR}$  &  Case (1)  &  Case (2)   & Case (3-1) & Case (3-2) \\
\hline 
 $g_{\pi \Delta^+ \Delta^+}$ & 0    &    -8.6  &  0    & -8.9  \\
 $g_{\pi \Delta^- \Delta^-}$ & 0    &     11   &  0    &  9.6  \\
 $g_{\pi \Delta^+ \Delta^-}$ & 5.2  &     1.9  &  0.25 &  0.81 \\
 $g_{\pi N^+ N^+}$           & 0    &     8.5  &  0    &  7.9  \\
 $g_{\pi N^- N^-}$           & 0    &    -7.5  &  0    & -7.5  \\
 $g_{\pi N^+ N^-}$           & -4.3 &    -1.7  & -0.21 & -0.73 \\
 $g_{\pi N^+ \Delta^+}$      & 0    &    -5.0  &  0    & -5.0  \\
 $g_{\pi N^+ \Delta^-}$      & 3.0  &     3.4  &  0.14 &  2.3  \\
 $g_{\pi \Delta^+ N^-}$      &-3.0  &     0.92 & -0.14 &  1.2  \\
 $g_{\pi N^- \Delta^-}$      & 0    &     5.3  &  0    &  5.1  \\
\hline \hline
\end{tabular}
\end{center}
\end{table}
Once the masses are determined, we obtain the one-pion coupling constants 
between two members of the quartet, which are shown in Table \ref{May2708tab1}. 
First, we consider qualitative features of the one-pion coupling constants.
It was found~\cite{Jido:1999hd} that in the case (1) the parity-non-changing 
interactions vanish, while the parity-changing interactions remain to be finite. However, even for the parity-changing 
interactions, their strengths are smaller than a typical order of one-pion 
interactions e.g. $g_{\pi NN } \sim 13$~\cite{Nagata:2003gg}. On the other hand, 
$g_{\pi RR}$ behaves in an opposite way in the case (2). All of the coupling 
constants survive in the case, where the parity-changing interactions are 
suppressed compared to the parity-non-changing ones. In addition, diagonal 
coupling constants are comparable to $g_{\pi NN }$, e.g. $g_{\pi \Delta_- \Delta_-}
= 11$. Interestingly, the cases 
(3-1) and (3-2) show different results, although they are the same assignment. 
This is caused by the difference of the ordering of the masses of the quartet, 
especially that of $\Delta(1920)$ and $N(2080)$. We turn back to this point later. 

Among various coupling constants, $g_{\pi \Delta(1232)\Delta(1232)}$ 
are investigated in several approaches. Quark models~\cite{Brown:1975di} and large 
$N_c$ ~\cite{Fettes:2000bb} predict large values, especially, 
$g_A^{\pi \Delta \Delta} = (9/5) g_A$ in large $N_c$ which gives $g_{\pi \Delta(1232) \Delta(1232)}\sim 30$. A light-cone QCD sum rule reported half of the quark model prediction~\cite{Zhu:2000zd} but still large values compared to our result. The $g_{\pi\Delta(1232)\Delta(1232)}$ were also determined in coupled channel analysis. Krehl et. al. obtained $g_{\pi\Delta\Delta}=31$~\cite{Krehl:1999km}, 
while Schneider et. al. obtained $g_{\pi \Delta \Delta}=12.5$~\cite{Schneider:2006bd}. 
In the case (1), $g_{\pi \Delta (1232) \Delta(1232)}$ vanishes, which is 
inconsistent with these studies. Krehl et. al. and Schneider et. al. also investigated $g_{\pi\Delta(1232)N(1520)}$ and obtained 
$g_{\pi N(1520) \Delta(1232)} = 0.95$ and 1.3, respectively. The present 
result $|g_{\pi \Delta(1232)N(1520)}| = 3.0$ is qualitatively consistent 
with these values. 

\begin{table}[htbp]
\begin{center}
\begin{minipage}{12cm}
\caption{Result for the $\pi N$ coupling constants and parameters. 
For the experimental data, see Table~\ref{May1910tab1} and \ref{May2510tab1}.}
\label{Feb2509tab1}
\end{minipage}
\begin{tabular}{ccccc}
\hline \hline 
  \multicolumn{5}{c}{$\pi N$ coupling constants Theo (Exp) [GeV$^{-1}$]  }\\
   & Case (1) & Case (2) & Case (3-1) & Case (3-2) \\
   \hline 
$\ds{\frac{g_{\pi N \Delta^+}}{\Lambda}}$ &  16  (15.7-16.0) & 7.2   (2.37-5.04) & 2.7   (0.825-2.33) & 1.8 (0.825-2.33) \\
$\ds{\frac{g_{\pi N \Delta^-}}{\Lambda}}$ &  14  (6.34-12.7) & 7.2   (3.23-6.20) & 8.9   (3.23-6.20)  & 12  (2.04-10.8)  \\
$\ds{\frac{g_{\pi N N^{* -}}}{\Lambda}}$  &  7.3 (7.64-9.30) & 4.2   (7.64-9.30) & 3.8   (4.63-5.23)  & 2.2 (4.45-7.84)  \\
$\ds{\frac{g_{\pi N N^{* +}}}{\Lambda}}$  &  1.3 (1.72-3.44) & -0.89 (1.72-3.44) & -0.44 (0.574-1.53) & 0.81(0.574-1.53) \\
\hline
$\chi_{\pi NR}^2$                  &  1.5 & 13  &  7.1   & 1.8 \\
\hline \hline
 \multicolumn{5}{c}{Parameters[GeV$^{-1}]$} \\ 
   & Case (1) & Case (2) & Case (3-1) & Case (3-2) \\  
   \hline 
$\ds{\frac{g_3 f_\pi}{\Lambda^2}}$ &  1.1  & -2.6  & -4.4  & -8.8 \\
$\ds{\frac{g_4 f_\pi}{\Lambda^2}}$ & -5.2  & -2.9  & -2.0  &  2.1 \\
$\ds{\frac{g_5}{\Lambda}}$         &  21   &  9.8  &  8.2  &  7.7 \\  
\hline \hline 
\end{tabular}
\end{center}
\end{table}
With regard to the $\pi N$ coupling constants $g_{\pi N R}$, we need to determine three 
parameters $g_3, g_4$ and $g_5$. Since $g_{\pi N R}$ are the functions of 
$g_3, g_4$ and $g_5$, we can determine them by $\chi^2$-minimum method with
$\chi^2_{\pi NR}= \sum_R (g_{\pi N R}-g_{\pi NR}^{{\rm (exp)}})^2/ (\delta g_{\pi N R}^{\rm (exp)})^2$. 
Here $g_{\pi NR}^{\rm (exp)}$ and $\delta g_{\pi NR}^{\rm (exp)}$ are the average and 
errors of the coupling constants determined from the experimental $\pi N$ decay widths. 
We obtain them by using a relation
$g_{\pi NR}^{\rm (exp)}/ \Lambda = \sqrt{\Gamma_{\pi N}^{(\rm exp)} /\tilde{\Gamma}_{\pi N}}$, where 
$\tilde{\Gamma}$ is $\pi N$ decay widths obtained by setting the coupling constant to be one, 
and $\Gamma_{\pi N}^{(\rm exp)}$ are the experimental values of the $\pi N$ decay widths shown in 
Table~\ref{May1910tab1} and \ref{May2510tab1}. The dimensional parameter $\Lambda$ does not play any 
role in the determination of the coupling constants because of the cancellation between the numerator and 
denominator in $\chi^2_{\pi NR}$. We obtain $\tilde{\Gamma}_{\pi N}$ by calculating the simplest tree diagram. 
Note that we can determine only absolute values of the coupling constants 
from the $\pi N$ decay widths. Hence, the positive sign of $g_{\pi NR}^{{\rm (exp)}}$ in 
Table~\ref{May1910tab1} and \ref{May2510tab1} are our assumption. The result is shown in Table~\ref{Feb2509tab1}.

The case (1) reproduces the reasonable values for the four $g_{\pi NR}$ with small $\chi^2_{\pi NR}$, 
which are almost within the ranges of the experimental values. In the case (2), $\chi^2_{\pi NR}$ 
value is significantly large. The discrepancy is mostly caused by the small values of the 
$\pi N$ decay width of $\Delta(1600)$ and $\Delta(1940)$. In the QS, it is favored that 
the average values of $g_{\pi NR}$ between $\Delta_\pm$ is larger than that between $N^*_\pm$, 
as is shown in Eq.~(\ref{Feb2509eq3}). Because of the same reason, $\chi^2_{\pi NR}$ is large for the case (3-1). 
We obtain reasonable results for the case (3-2) with small $\chi^2_{\pi NR}$. 
Our result underestimates the value of $g_{\pi NR}$ for $R=N(2080) (N^*_-)$, which gives 
$\pi N$ decay widths half of the minimum of the experimental values. 

\subsection*{Mass pattern and one-pion coupling constant}

The quartet scheme shows two different behavior for the one-pion coupling constants, 
as shown in Table~\ref{May2708tab1}. Especially, the assignment 
$(\Delta(1920)$, $\Delta(1940)$, $N(2080)$, $N(1900)$) shows two different 
behavior, depending of the choice of the experimental data.
Equations~(\ref{Nov0209eq1}) shows that the one-pion coupling constants 
are controlled by the mixing angles. The cases (1) and (3-1) correspond to the
maximally mixing with the angles $\theta_{N, \Delta}=45^\circ$, 
while the cases (2) and (3-2) correspond to moderate mixing. 
Since the mixing angles are the functions of $m_0$  and $(g_1-g_2)f_\pi$ as shown in 
Eqs.~(\ref{Jul0508eq1}) and (\ref{Jul0508eq2}), we can understand
the behavior of the one-pion coupling constants, comparing
$m_0$ with $(g_1-g_2)f_\pi$. These parameters also determine the masses of the quartet. 
Therefore, we can relate the masses to the one-pion constants.

\begin{figure}[htbp]
\begin{center}
\includegraphics[width=8cm]{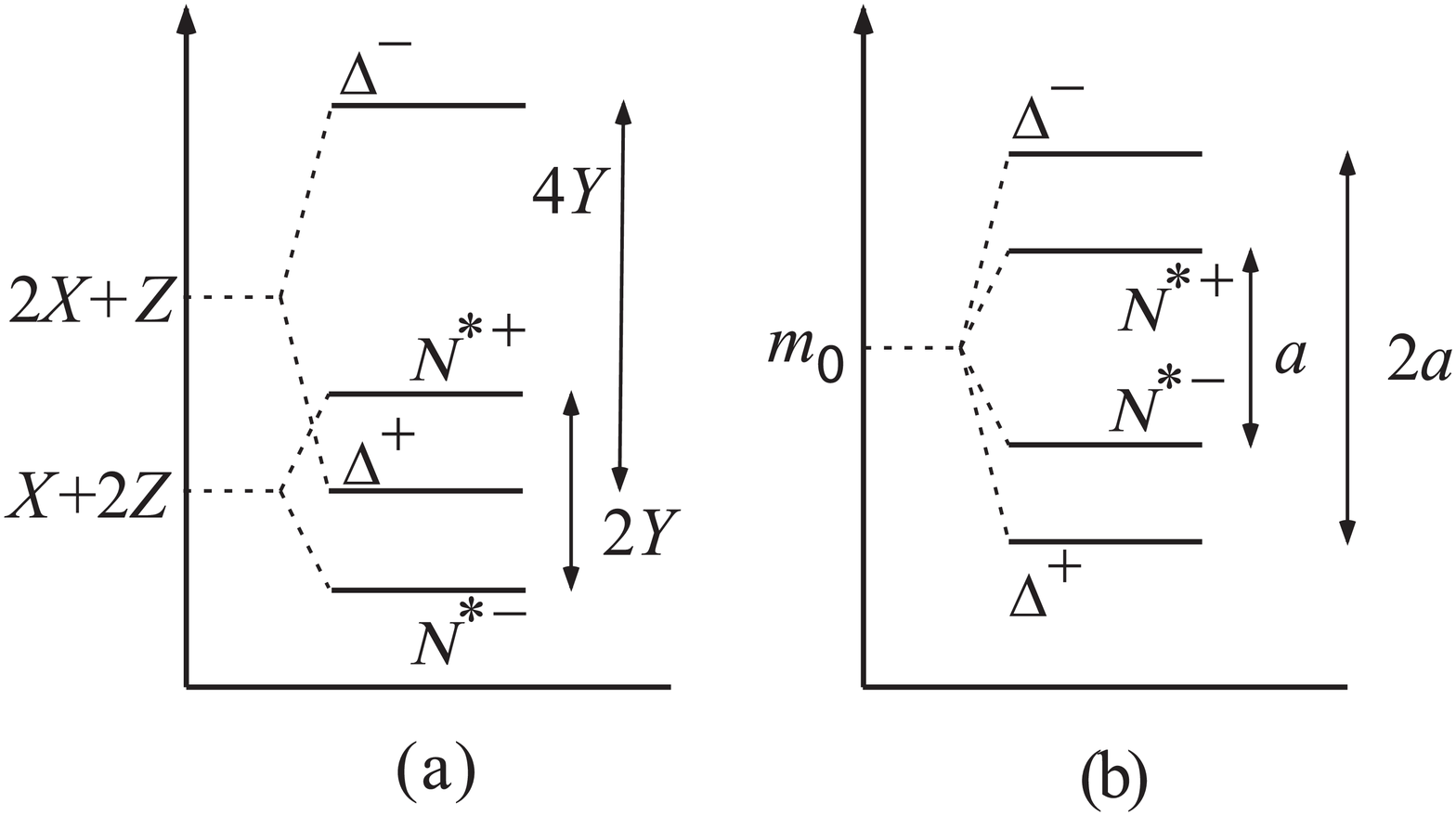}
\begin{minipage}{10cm}
\caption{Schematic figures for the mass pattern of the QS. 
(a) small $m_0$ case.  (b)  $m_0$-dominant case. 
}\label{Mar1310fig1}
\vspace{0.2cm}
\end{minipage}
\end{center}
\end{figure}

In order to understand their relation,  we approximate the masses in two ways.
In the small $m_0$ case, the masses are, up to ${\cal O}(m_0^2)$, given by  
\begin{align}
m_{\Delta^\pm} & = 2 X \mp 2 Y + Z,  \nn\\
m_{N^\pm}      & =   X \pm   Y + 2Z, \nn
\end{align}
where $X= f_\pi |g_1-g_2| /4$, $Y= (g_1+g_2)f_\pi /4$ and $Z= 
4m_0^2/ (f_\pi |g_1-g_2|)$.  
In the $m_0$ dominant case, they are, up to ${\cal O}( \left(f_\pi/m_0\right))$, given  by 
\begin{align}
m_{\Delta^\pm} & = m_0 \mp 2 a  ,\nn \\
m_{N^{*\pm}}   & = m_0 \pm a, \nn
\end{align}
where  $a=(g_1+g_2)f_\pi / 4$.
The mass patterns for these cases are shown in Fig.~\ref{Mar1310fig1}.
The two cases are different in the ordering of $\Delta^+$ and $N^{*-}$. 
In the $m_0\to0$ limit, they have mass ratio $2:1$ and $\Delta^+$ is heavier than 
$N^{*-}$. Small values of $m_0$ do not change this ordering, which 
corresponds to the left panel in Fig.~\ref{Mar1310fig1}. When $m_0$ becomes much larger, 
the ordering is changed and $\Delta^+$ becomes lowest-state. 
The cases (1) and (3-1) correspond to the mass pattern shown in the right panel 
in Fig.~\ref{Mar1310fig1}, while the cases (2) and 
(3-2) correspond to the left panel. Actually, $m_0$ is not small 
in the cases (2) and (3-2), but comparable to $(g_1-g_2) f_\pi$. 
However, the left panel in Fig.~\ref{Mar1310fig1} well described the mass pattern of 
these cases. 
Using Eqs.~(\ref{Jul0508eq1}) and (\ref{Jul0508eq2}), mixing angles in the small 
$m_0$ case takes moderate values and all the one-pion coupling constants survive.
On the other hand  in the $m_0$-dominant case, mixing angles are $\theta_{N, \Delta}\sim \pi / 4$
and the parity-non-changing interactions vanish.
Thus, the behavior of the one-pion coupling constants is related 
to the mass pattern of the quartet. According to this discussion, 
the cases (3-1) and (3-2) are different due to the ordering of $\Delta(1920)$ and $N(2080)$, 
although they describe the same assignments. This is the reason why the assignment 
$(\Delta(1920)$, $\Delta(1940)$, $N(2080)$, $N(1900)$) is sensitive to the choice of the 
experimental data. 
This discussion can be applied to other assignments we do not take into account. 
As we have mentioned, other three assignments reproduces the masses of the quartet 
with $\chi^2_{\rm mass}$ less than one :($\Delta(1600)$, $\Delta(1700)$, $N(1700)$, $N(1720)$), 
($\Delta(1600)$, $\Delta(1940)$, $N(1700)$, $N(1900)$), 
($\Delta(1920)$, $\Delta(1940)$, $N(1700)$, $N(1720)$). 
According to the above discussions, the first and second cases correspond to 
maximally-mixing with the vanishing of the parity-non-changing interactions, 
while all the coupling constants survive in the third case. 

\section{Summary}
We have investigated the possibility that chiral partners exist in spin-$\thalf$
baryon sector by considering the quartet scheme, where four spin-$\thalf$ baryons, 
$P_{33}$, $D_{33}$, $D_{13}$ and $P_{13}$, form the chiral multiplets 
$(1,\half)\oplus (\half,1)$ with the mirror assignment. Using the 
$SU(2)_R\times SU(2)_L$ Lagrangian, we tried to find a set of four 
baryons suitable for the chiral quartet. We discussed three assignments: 
(1) $(\Delta(1232), \Delta(1700), N(1520), N(1720))$, 
(2)  $(\Delta(1600), \Delta(1940), N(1520), N(1720))$, 
(3-1) and (3-2) $(\Delta(1920)$, $\Delta(1940)$, $N(2080)$, $N(1900)$). 
Here we investigated $(\Delta(1920)$, $\Delta(1940)$, $N(2080)$, $N(1900)$) using 
two data sets.

For the case (1) we found that there is significant discrepancy for the masses, 
which implies the mass of $\Delta(1232)$ is too small for the quartet scheme. 
In addition, the vanishing of $g_{\pi \Delta(1232) \Delta(1232)}$ inconsistent with 
other theories. Considering the discrepancy for the masses and the inconsistencies of 
$g_{\pi \Delta(1232) \Delta(1232)}$, it seems that this case is less suitable for 
the quartet. 

For the case (2), the masses of the observed baryons agree well with the mass pattern 
of the QS. Among all the possible assignments, the $\chi^2$ value becomes the smallest 
in this case. Considering the masses, this case is most suitable for the quartet. 
Regarding the $\pi N$ interactions, this case does not reproduce reasonable results. 

For the assignment $(\Delta(1920)$, $\Delta(1940)$, $N(2080)$, $N(1900)$), we 
consider two cases (3-1) and (3-2) with the use of different data sets because of the 
variety of the experimental data. Both cases reproduce the masses of the quartet with 
$\chi^2$ less than one. The one-pion coupling constants for this assignment are quite 
sensitive to the ordering of the masses of $\Delta(1920)$ and $N(2080)$. If the mass 
of $\Delta(1920)$ is smaller than that of $N(2080)$, only the parity-changing one-pion 
interactions survive. On the other hand, if the mass of $N(2080)$ is smaller, all 
the coupling constants are finite and the parity-non-changing interactions are larger 
than the parity-changing ones. Regarding the $\pi N$ interactions, we obtained reasonable 
results for the case (3-2).

For further confirmation, experiments or lattice calculations for the one-pion 
coupling constants are needed. For instance, we can test the validity of the case 
(2) using  coupling constants such as $g_{\pi N(1520)N(1520)}$, $g_{\pi N(1720)N(1720)}$ 
and $g_{\pi N(1520)N(1720)}$. For the further study of the assignment 
($\Delta(1920)$, $\Delta(1940)$, $N(2080)$, $N(1900)$), we need information 
about the masses because of a variety of the data. Especially, detailed 
information of the masses of $\Delta(1920)$ and $N(2080)$ are needed, because the 
one-pion coupling constants are sensitive to the ordering of the masses of them. 
If the mass ordering are determined, we can test this assignment using one-pion 
coupling constants such as $g_{\pi \Delta(1920)\Delta(1920)}$. 

It is important to extend the present framework with the inclusion of 
higher-dimensional chiral representations for the nucleon. For the $\pi N$ 
interactions with the quartet, we adopted the assumption that the nucleon 
belongs to the fundamental chiral representation. There are other possibilities
for the nucleon's chiral representation. Hence, the disagreements for the $\pi N$ 
interactions may come from this assumption and can be resolved by including 
higher-dimensional chiral representations for the nucleon. Furthermore, it 
may be possible to test the nucleon's chiral representations through the 
$\pi N$ interactions with the quartet, if we can confirm the QS by using 
the one-pion interactions for the quartet. 

In the present study, we employed the effective Lagrangian approach, where we 
truncated higher-order terms in the Lagrangian and we neglected quantum effects. 
With the high-lying baryons in the multiplet, we need to include various resonances 
in order to evaluate the quantum effects properly, which would cause additional 
difficulties. Rather, it is desired to reproduce and confirm the present result 
using different method. For instance, an algebraic method proposed by Weinberg 
is one of the useful method to study chiral partners. This method is based on 
the commutation relations derived from the superconvergence property of pion-nucleon 
scattering amplitudes, and can be applied to baryons~\cite{Weinberg:1969hw,Beane:2002ud,Hosaka:2001ti}. 
We have already started a study along this line in Ref.~\cite{Nagata:2008cq}. 
\section*{Acknowledgments}
This work was partly supported by Grant-in-Aid for 
Scientific Research on Innovative Areas(20105003) and
Grant-in-Aid for Scientific Research (B20340055).

\appendix

\section{Fierz Transformation}
\label{Oct0209sec1}
We show the derivation of Eqs.~(\ref{Sep2709eq1}).
We define totally anti-symmetric fields as linear combinations of
Eqs.~(\ref{eq:22feb2008eq1})
\begin{subequations}
\begin{align}
B_N   &=\inner{a_N}{\phi_N}, \\
B_\Delta &=\inner{a_\Delta}{\phi_\Delta}, 
\end{align}
where 
\begin{align}
\vec{\phi}_{N} &=(N_{V}^\mu,  N_{A}^\mu, N_{T}^\mu),\\
\vec{\phi}_{\Delta} &=(\Delta_{A}^{\mu i}, \Delta_{T}^{\mu i}),\\
\vec{a_N} &=(a_1^N,a_2^N,a_3^N),\\
\vec{a_\Delta} &=(a_1^\Delta,a_2^\Delta).
\end{align}
\end{subequations}
The coefficients $\vec{a}_N$ and $\vec{a}_\Delta$ are determined by the 
totally anti-symmetric condition, which is implemented by 
the anti-symmetric condition under the interchange between the second and 
third quark is given by 
\begin{align}
{\cal F} [B_{{\rm n}}]=-[B_{{\rm n}}], ({\rm n}=N,\Delta),
\label{eq:24feb2008eq1}
\end{align}
where ${\cal F}[B]$ denotes a baryon field obtained from the Fierz 
transformation of $B$. Fierz transformation formula is given in Ref.~\cite{Nagata:2007di}. 
This equation can be read  as two kinds of the eigen-value problems
: (a) for the vector space $\vec{B}_{N,\Delta}$,  and (b) for the vector space $\vec{a}_{N,\Delta}$.
The eigen-value problem (a) gives identities between the baryon operators
\begin{subequations}
\begin{align}
N_{V}^\mu &=N_{A}^\mu, 2 N_{A}^\mu=N_{T}^\mu,\\
\Delta_{A}^{\mu i} &= -\Delta_{T}^{\mu i},
\end{align}%
\label{Apr1008eq2}%
\end{subequations}%
which reduce the number of the independent fields~\cite{Nagata:2007di,Ioffe:1981kw,Chung:1981cc,Espriu:1983hu}.
The eigen-value problem (b) determines the values of the coefficients 
$\vec{a}_N$ and $\vec{a}_\Delta$
\begin{subequations}
\begin{align}
\vec{a}_N & =(3,1,1),\\
\vec{a}_\Delta & =(-2, 1),
\end{align}%
\label{Apr1008eq3}%
\end{subequations}%
with which $B_N$ and $B_\Delta$ are totally anti-symmetric. 
This determine the ratio between $N_V^\mu$ and $N_A^\mu$ in $N_1^\mu$.
It is convenient to replace $N_T^\mu$ by $N_V^\mu $ and $N_A^\mu $
and $\Delta_T^{\mu i}$ by  $\Delta_A^{\mu i}$ with the use of Eqs.~(\ref{Apr1008eq2}), which can be done 
without the change of chiral transformation properties of $B_N$ and $B_\Delta$.
Finally, we obtain Eqs.~(\ref{Sep2709eq1}).

\section{Alternative derivation of chiral properties}
\label{Feb0110sec1}

We show an alternative derivation of  the chiral transformation properties of
$(1,\half)\oplus (\half, 1)$ and the mass relation. 
Starting point is a standard definition of the transformation in terms of  the chiral algebra 
between charges and fields. 
In general, the $SU(2)_A$ transformation is given by $\psi^\prime = \psi + i a^i [Q^i_A, \psi]$
with generators  $Q_A^i, (i = 1, 2, 3)$  and infinitesimal parameters $a^i$ for the $SU(2)_A$ transformation.
We describe $(1,\half)\oplus (\half,1)$ by product of 
the isovector and isospinor $\psi^i = (\psi^i)_a, (a= 1, 2)$. For simplicity, we suppress the Lorentz indices in this section.

In the left- and right-handed representation, they correspond to
$\psi^i_R = (1, \half)$ and $\psi^i_L = (\half, 1)$ : $\psi^i_R = (1, \half)$ transforms as $I=1$ under $SU(2)_R$ and $I=\half$ under $SU(2)_L$, while  $\psi^i_L = (\half, 1)$
transforms $I=\half$ under $SU(2)_R$ and $I=1$ under $SU(2)_L$. Note that this field 
$\psi^i$ corresponds to $\Delta_T^i$ and $N_T$ in Eq.~(\ref{eq:22feb2008eq2}). 
It is easy to check that $N_A$, $N_V$ and  $\Delta_A$ consist of  $(RL) R$, $(RL) L$,  $(LR) R$ and  $(LR) L$, 
while $N_T$ and $\Delta_T$ contain $(RR) L$ and $(LL)R$. Jido et. al. employed 
$(RR) L$ and $(LL)R$ for the description of $(1,\half)\oplus (\half, 1)$~\cite{Jido:1999hd}.
The chiral transformations of these fields are given by
\begin{align}
\left\{ \begin{array}{l}
\delta_R^a \psi_{R i}^b = \epsilon^{abc} (\psi_r)_i^c , \\
\delta_R^a \psi_{L i}^b = i t^a \psi_l^b,
\end{array}\right.
\left\{ \begin{array}{l}
\delta_L^a \psi_{R i}^b = \epsilon^{abc}  (\psi_r)_i^c , \\
\delta_L^a \psi_{L i}^b = i  t^a \psi_r^b,
\end{array}\right.
\end{align}
where we have defined $\delta^a \psi^b = -i [Q^a, \psi^b]$. 
Using $Q_V^a = Q_R^a + Q_L^a$ and $Q_A^a= Q_R^a-Q_L^a$, we obtain 
$SU(2)_V$ and $SU(2)_A$  transformation properties 
\begin{align}
\delta^a_V  \psi_i^b &= \left[  (\epsilon^{abc} + i t^a \delta^{bc}) \right] \psi^c, \\
\delta^a_A \psi^b    &=  \gamma_5 (\epsilon^{abc} -i t^a \delta^{bc} ) \psi_i^c.
\end{align}
Employing an isospurion formalism, $I=\half$ and $I=\thalf$ components are obtained by 
$\psi_{1/2}=\tau^i \psi^i$ and $\psi_{3/2}^i = P^{ij}_{3/2} \psi^j$.
After the irreducible decomposition, we obtain 
\begin{subequations}
\begin{align}
\delta_A^a \psi_{1/2}   &= \half i\gamma_5 \left[ \frac{5}{3} \tau^a \psi_{1/2} - 4 \psi_{3/2}^a\right], \\
\delta_A^a \psi_{3/2}^b &= \half i\gamma_5 \left[ \tau^a \psi_{3/2}^b -\frac{2}{3} \tau^b \psi_{3/2}^a - \frac{4}{3} P^{ba}_{3/2} \psi_{1/2}\right].
\end{align}%
\label{Mar1410eq1}%
\end{subequations}%
Here note that  the coefficients differ from Eqs.~(\ref{Jul0608eq1}). This is because $\psi_{1/2}$ and 
$\psi_{3/2}^a$ describe $N_T$ and $\Delta_T^i$, respectively.
Using Eqs.~(\ref{Sep2709eq1}) and (\ref{Apr1008eq2}),  we obtain $\psi_{1/2} = N_T = 2\sqrt{3} N_1$ 
and $\psi_{3/2}=\Delta_T = -2 \Delta_1$. Substituting these relations into Eqs.~(\ref{Mar1410eq1}), 
we reproduce Eqs.~(\ref{Jul0608eq1}).

Considering  the $I_z=\half$ components, it is easy to show that the 
$SU(2)_A$ transformations of the $I=\half$ and $\thalf$ fields
\begin{align}
\delta_A^a\left( \begin{array}{c} 
\psi_{1/2}^{I_z=\half} \\ 
\psi_{3/2}^{I_z=\half}
\end{array} \right)
= T
\left( \begin{array}{c} 
\psi_{1/2}^{I_z=\half} \\ 
\psi_{3/2}^{I_z=\half}
\end{array} \right),   \;\;
T=\half\left( \begin{array}{cc}
\frac{5}{3} & \frac{4\sqrt{2}}{3} \\
\frac{4\sqrt{2}}{3} & \frac{1}{3}
\end{array}
\right),
\end{align}
where $T$ is  the axial-transformation matrix Eq.~(\ref{Mar1410eq1}) for $I_z=\half$ components. 
We introduce the mass matrix for $(\psi_{1/2}^{I_z=\half}, \psi_{3/2}^{I_z=\half})^T$ as 
$M = \diag(a, b)$ with $a$ and $b$ being the masses of $\psi_{1/2}$ and $\psi_{3/2}$. We also introduce 
the pion interaction matrix $M_{\pi}$ for their pseudo-scalar couplings.
With chiral invariance,  the matrices $T$, $M$ and $M_\pi$ must obey 
\begin{align}
M &=  \{T, M_\pi\}, \nn\\
M_\pi &= \{ T, M\}, \nn
\end{align}
which leads to  a double-commutation relation
\begin{align}
M = \{ T, \{ T, M\} \}.
\label{Mar1410eq2}
\end{align}
This double-commutation relation gives  $a = -2 b$, which reproduces 
the mass relation between $N_1^\mu$ and $\Delta_1^{\mu i}$. 
Note that the double commutator Eq.~(\ref{Mar1410eq2}) is the necessity condition 
of chiral invariance.

\baselineskip 5mm

\end{document}